\documentclass[12pt,a4paper]{article}

\usepackage{amsgen,amstext,amsbsy,amsopn, bm, enumerate, psfrag, epsf, enumerate, mathtools, comment}
\usepackage[colorlinks=true,citecolor=blue]{hyperref}
\usepackage[margin=1in]{geometry}
\usepackage{titlesec}
\newcommand{\bbS}{\bar{\bm{S}}}
\newcommand{\rowbind}{\raisebox{.5pt}{\textcircled{\raisebox{-.9pt} {r}} } }

\usepackage{kz}

\def\spacingset#1{\renewcommand{\baselinestretch}%
{#1}\small\normalsize} \spacingset{1}

\spacingset{1.47} % DON'T change the spacing!

\providecommand{\keywords}[1]
{
  \small	
  \textbf{\textit{Keywords---}} #1
}

\title{BEAUTY Powered BEAST}
\author{
Kai Zhang
  \footnote{Kai Zhang is Associate Professor (E-mail: zhangk@email.unc.edu), Department of Statistics and Operations Research, University of North Carolina at Chapel Hill, Chapel Hill, NC 27599.}
  \and Wan Zhang
  	\footnote{Wan Zhang is Assistant Professor (E-mail: wanzhang@amss.ac.cn), The Academy of Mathematics and System Science Center for Forecasting Science, Chinese Academy of Sciences, Beijing, China.} 
  	\and Zhigen Zhao
  	\footnote{Zhigen Zhao is Associate Professor (E-mail: zhaozhg@temple.edu), Department of Statistical Science, Temple University, Philadelphia, PA 19121.} 
  	\and Wen Zhou
  	\footnote{Wen Zhou is Associate Professor (E-mail: w.zhou@nyu.edu), Department of Biostatistics, School of Global Public Health, New York University, New York, NY 10003.} 
}
\date{\vspace{-3em}}

\begin{document}
\maketitle

%%%%%%%%%%%%%%%%%%%%%%%%%%%%%%%%%%%%%%%%%%%%%%%%%%%%%%%%%%%%%%%%%%%%%%%%%%%%%%

\begin{abstract}
We study distribution-free goodness-of-fit tests with the proposed {\bf B}inary {\bf E}xpansion {\bf A}pproximation of {\bf U}niformi{\bf TY} (BEAUTY) approach. This method generalizes the renowned Euler's formula, and approximates the characteristic function of any copula through a linear combination of expectations of binary interactions from marginal binary expansions. This novel theory enables a unification of many important tests of independence via approximations from specific quadratic forms of symmetry statistics, where the deterministic weight matrix characterizes the power properties of each test. To achieve a robust power, we examine test statistics with data-adaptive weights, referred to as the {\bf B}inary {\bf E}xpansion {\bf A}daptive {\bf S}ymmetry {\bf T}est (BEAST). {For any given alternative, we demonstrate that the Neyman-Pearson test can be approximated by an oracle weighted sum of symmetry statistics.} The BEAST with this oracle provides a useful benchmark of feasible power. To approach this oracle power, we devise the BEAST through a regularized resampling approximation of the oracle test. The BEAST improves the empirical power of many existing tests against a wide spectrum of common alternatives and delivers a clear interpretation of dependency forms when significant. 
\end{abstract}

\vspace{-1.1em}
\keywords{Nonparametric Inference; Goodness-of-fit Test; Test of Independence; Resampling; Characteristic Function; Euler's Formula}

\section{Introduction}\label{sec: intro}

As we enter  the era of data dominance,  the prevalence of complex datasets presents significant challenges to traditional parametric inference methods, which often lose efficacy in practical applications, as flawless models are rarely attainable through scientific theories alone. In contrast, nonparametric methods deliver more robust inference, rendering them increasingly appealing for practical use. Hypotheses addressing particular structures of distributions are classified as goodness-of-fit tests \citep{lehmann2006testing}. Notable progress within this domain is extensively documented in the literature, as exemplified by moment or cumulant based methods \citep{anderson1954test,stephens1976asymptotic}, likelihood ratio based methods \citep{cressie1984multinomial,fan2001goodness,zhang2002powerful}, empirical process based methods \citep{genest2006goodness,escanciano2006goodness,jager2007goodness,genest2009goodness,kaiser2012goodness}, kernel based methods \citep{gonzalez2013updated,sen2014testing}, and randomization based methods \citep{jankova2020goodness,kim2020dimension,barber2022testing}, along with the references cited therein.

Without loss of generality, we consider a $p$-dimensional distribution in $[-1,1]^p$ for notation convenience. Let $\bU=({U_1,\ldots,U_p})^T$ denote a $p$-dimensional vector whose marginal distributions are continuous and whose joint distribution $\Pb_{\bU}$ has a support within $[-1,1]^p$. With the above notation, the goodness-of-fit test for some hypothesized distribution $\Pb$ can be written as follows: 
\begin{equation}\label{eq: test_gof}
	H_0: \Pb_{\bU} = \Pb ~~v.s.~~ H_1: \text{Dist}(\Pb_{\bU},\Pb) \ge \delta
\end{equation}
for some distance $\text{Dist}(\cdot, \cdot)$ between distributions and some $0<\delta\le 1.$ Some common choices of $\text{Dist}(\cdot, \cdot)$ include the total variation (TV) distance $\text{TV}(\cdot,\cdot)$ \citep{zhang2019bet} and the $L_2$ distance \citep{berrett2021optimal}. 

Two important special cases of goodness-of-fit tests are the test of uniformity and the test of independence. The test of uniformity can be formulated as 
\begin{equation}\label{eq: test_unif}
H_0: \Pb_{\bU} = \Pb_0 ~~v.s.~~ H_1: \text{Dist}(\Pb_{\bU},\Pb_0) \ge \delta,
\end{equation}
where $\Pb_0=\text{Unif}[-1,1]^p$ denotes the uniform distribution. The test of independence can be formulated as 
\begin{equation}\label{eq: test_ind}
H_0: \Pb_{(\bU_1,\bU_2)} = \Pb_1\times \Pb_2 ~~v.s.~~ H_1: \text{Dist}(\Pb_{(\bU_1,\bU_2)},\Pb_1\times \Pb_2) \ge \delta,
\end{equation}
where $\bU_1$ and $\bU_2$ are $p_1$ and $p_2$ dimensional random vectors with distributions $\Pb_1$ and $\Pb_2$ respectively. Applications of the test of uniformity include \cite{rivero2020flow,liang2001testing}. Important developments in distribution-free tests of independence include cumulative distribution function (CDF) based methods \citep{hoeffding1948non,blum1961,genest2005locally,kojadinovic2009tests,genest2019testing,chatterjee2019new,cao2020correlations}, kernel based methods \citep{szekely2007,gretton2007kernel,zheng2012generalized,sejdinovic2013,pfister2016kernel,Zhu2017,jinmatteson18,Balakrishnan2019hypothesis,shi2020rate,deb2020measuring,Geenens2021new,berrett2021optimal,berrett2021usp}, binning based methods \citep{miller1982maximally,reshef2011detecting,heller2012consistent,kinney2014equitability, heller2016consistent, heller2016multivariate,ma2019fisher,lee2023beret}, and references therein.

To facilitate the analysis of large datasets, some desirable attributes of distribution-free tests of independence include (a) a robust high power against a wide range of alternatives, (b) a clear interpretation of the form of dependency upon rejection, and (c) a computationally efficient algorithm. An example of recent development towards these goals is the binary expansion testing (BET) framework and the Max BET procedure in \cite{zhang2019bet}. It was shown that the Max BET is minimax optimal in power under mild conditions, has clear interpretability of statistical significance and is implemented through computationally efficient bitwise operations \cite{zhao2023fast}. Potential improvements of the Max BET include the followings: (a) The procedure is only univariate and needs to be generalized to higher dimensions. (b) The multiplicity correction is through the conservative Bonferrnoni procedure, which leaves room for further enhancement of power. In \cite{lee2023beret}, random projections are applied to the multivariate observations to reduce the dimension to one, so that the univariate methods of \cite{zhang2019bet} are applicable. An ensembled approach involving distance correlation is further used to improve the power towards monotone relationships.

In this paper, we develop an in-depth understanding of the BET framework and construct a class of powerful distribution-free goodness-of-fit tests, encompassing both the uniformity test and the independence test between a random vector and a random variable.  While many tests have been devised for this problem, \cite{zhang2019bet} showed that uniform consistency for testing ~\eqref{eq: test_ind} is generally unachievable for all alternative distributions. Analogous results for the $L^2$-distance was recently documented by \cite{berrett2021optimal}. These results reaffirm earlier observations by \cite{lecam1973convergence, barron1989uniformly, wasserman:2019}.

Practically, this finding indicates that each test encounters a ``blind spot,'' resulting in a significant loss of power. To mitigate the power loss arising from non-uniform consistency, \cite{zhang2019bet}  introduced the binary expansion statistics (BEStat) framework to restrict the space of alternative distributions up to a suitable finite resolution. The BEStat approach is inspired by the classical probability result of the binary expansion of a uniformly distributed random variable \citep{kac1959statistical}, as stated below.  
\begin{theorem}\label{thm: be}
	If $U \sim \mathrm{Unif}[-1,1],$ then $U=\sum_{d=1}^\infty 2^{-d}{A_d}$ where $A_d \stackrel{i.i.d.}{\sim} \text{Rademacher}$, that is $A_d\in\{-1,1\}$ with equal probabilities.
\end{theorem} 
Theorem~\ref{thm: be} allows the approximation of the $\sigma$-field generated by $U$ using $U_D=\sum_{d=1}^{D} {2^{-d}}{A_d}$ for any positive integer depth $D$. This filtration approach for testing uniformity facilitates a universal distribution approximation, an identifiable model, and uniformly consistent tests at any depth $D$. The testing framework based on the binary expansion filtration approximation is referred to as the binary expansion testing (BET). Specifically, the BET of approximate uniformity for $\bU_D=(U_{1,D},\ldots,U_{p,D})^T$ is   
\begin{equation}\label{eq: test_ind_unif_D}
	H_{0,D}: \Pb_{\bU_D} = \Pb_{0,D} ~~v.s.~~ H_{1,D}: \text{Dist}(\Pb_{\bU_D},\Pb_{0,D}) \ge \delta,
\end{equation}
where $\Pb_{0,D}$ is the uniform distribution over $p$-dimensional dyadic rationals $\{2^{-D}(1-2^D)+2^{-D+1}k,k=0,1,\ldots,2^{D}-1 \}^p.$

\subsection{Our Contributions}

Our study of the BET framework is inspired by the celebrated Euler's formula,
\begin{equation*}\label{eq: euler}
	e^{ix}=\cos{x}+i\sin{x},~~~~~~\forall x \in \RR,
\end{equation*}
which is often regarded as one of the most beautiful equations in mathematics. In particular, when $x=\pi$, one has Euler's identity, $e^{i\pi}+1=0$, which connects the five most important numbers in mathematics $0,1, i,e,\pi$ in one simple yet deep equation. Beside the beauty of this equation, how is it useful for statisticians? To see that, consider any binary variable $A$ (not necessarily symmetric) which takes values $-1$ or $1$. Through the parity of the sine and cosine functions, one can easily show the following binary Euler's equation. Since we were not aware of any reference of this equation in literature, we formally state it below.
\begin{lemma}[Binary Euler's Equation]\label{thm: bee}
	For any binary random variable $A$ with possible outcomes of $-1$ or $1$, it holds that for any $x \in \RR,$
	\begin{equation}\label{eq: bee}
		e^{iA x}=\cos{x}+iA\sin{x}.
	\end{equation}
\end{lemma}
Lemma~\ref{thm: bee} generalizes Euler's formula with additional randomness from a binary variable and reduces its complex exponentiation to its linear polynomial. To the best of our knowledge, no other random variables enjoy the same remarkable attribute. Moreover, note that the random variable $e^{i A x}$ in \eqref{eq: bee} is closely related to characteristic functions, particularly when it is combined with the binary expansion in Theorem~\ref{thm: be}. For example, for $U=\sum_{d=1}^\infty 2^{-d}{A_d} \sim \text{Unif}[-1,1]$ and for any $t \in \RR,$ we have
\begin{equation}\label{eq: ber}
	e^{i U t}= e^{it \sum_{d=1}^{\infty} \frac{A_d}{2^d}}= \prod\nolimits_{d=1}^{\infty} e^{ \frac{iA_d t}{2^d} } = \prod\nolimits_{d=1}^{\infty} 
	\{\cos{(t/2^d)}+iA_d \sin{(t/2^d)}\}.
\end{equation}
The complex exponent of $U$ can be approximated by a polynomial of the binary variables in its binary expansion! Moreover, we show in Section~\ref{sec: beauty} that this approximation is universal for {\it any} $p$-dimensional vector supported within $[-1,1]^p$. We refer this universal binary interaction approximation of the complex exponent and the characteristic function as the {\bf B}inary {\bf E}xpansion {\bf A}pproximation of {\bf U}niformi{\bf TY} (BEAUTY) in Theorem~\ref{thm: beauty}.

Based on the BEAUTY, in this paper we make the following three main novel contributions to the problem of nonparametric tests of independence:

{\it 1. A unification of important nonparamatric tests of independence.} In Section~\ref{sec: unification}, we show that many important tests of independence in literature can be approximated by some quadratic forms of symmetry statistics, which are shown to be complete sufficient statistics for dependence in \cite{zhang2019bet}. In particular, each of these test statistics corresponds to a different deterministic weight matrix in the quadratic form, which in turn dictates the power properties of the test. Therefore, this deterministic weight in existing test statistics creates the key issue on uniformity and robustness of the test, as it may favor certain alternatives but cause a substantial loss of power for other alternatives. Following this observation, we consider a test statistic that has data-adaptive weights to make automatic adjustments under different situations so as to achieve a robust power. We refer this test as the {\bf B}inary {\bf E}xpansion {\bf A}daptive {\bf S}ymmetry {\bf T}est (BEAST), as described in Section~\ref{sec: beast}. 

{\it 2. A benchmark of feasible power from the BEAST with oracle.} We begin by considering the test of uniformity. By utilizing the properties of the binary expansion filtration, we show in a heuristic asymptotic study of the BEAST a surprising fact that for any given alternative, the Neyman-Pearson test for testing uniformity can be approximated by a weighted sum of symmetry statistics. We thus develop the BEAST through an oracle approach over this Neyman-Pearson one-dimensional projection of symmetry statistics, which quantifies a boundary of feasible power performance. Numerical studies in Section~\ref{sec: simu} show that the BEAST with oracle leads a wide range of prevailing tests by a surprisingly huge margin under all alternatives we considered. This enormous margin thus provides helpful information about the potential of substantial power improvement for each alternative. To the best of our knowledge, there is no other type of similar approach or results to study the potential performance of a test of uniformity or independence. Therefore, the BEAST with oracle sets a novel and useful benchmark for the feasible power under any alternative. Moreover, it provides guidance for choosing suitable weights to boost the power of the test.

{\it 3. A powerful and robust BEAST from a regularized resampling approximation of the oracle.} Motivated by the form of the BEAST with oracle, we construct the practical BEAST to approximate the optimal power by approximating the oracle weights in testing uniformity. The proposed BEAST combines the ideas of resampling and regularization to obtain data-adaptive weights that adjusts the statistic towards the oracle under each alternative. Here resampling helps the approximation of the sampling distribution of the oracle test statistic, and regularization screens the noise in the estimation of optimal weights. This test is applied to the problem to the test of independence. Simulation studies in Section~\ref{sec: simu} demonstrate that the BEAST improves the power of many existing tests of univariate or multivariate independence against many common forms of non-uniformity, particularly multimodal and nonlinear ones. Besides its robust power, the BEAST provides clear and meaningful interpretations of statistical significance, which we demonstrate in Section~\ref{sec: data}. We conclude our paper with discussions in Section~\ref{sec: disc}. Details of notation, theoretical proofs and additional numerical results are deferred to Supplementary materials.

\section{The BEAUTY Equation}\label{sec: beauty}
To further understand the BET framework, we first develop the general binary expansion for any random vector supported within $[-1,1]^p$ as in Lemma \ref{lem: binrary:approx}.

\begin{lemma}\label{lem: binrary:approx}
	Let $\bU=(U_1,U_2,\cdots, U_p)^T$ be a random vector supported within $[-1,1]^p$. There exists a sequence of random variables $\{A_{j,d}\}$, $j=1,2,\cdots,p$, $d=1,2,\cdots,D$, which only take values $-1$ and $1$, such that $
	\max_{1\le j\le p}\{ |U_j  - U_{j,D}|\} \to 0$ uniformly as $D \rightarrow \infty$, where $
	U_{j,D} = \sum_{d=1}^D \left(A_{j,d}\right)/2^d.$
\end{lemma}

A classical construction of $A_{j,d}$'s is to consider the binary numeral system representation of real numbers, or data bits \citep{kac1959statistical,zhang2019bet}. We refer the collection of variables $\{A_{j,d}\}$ as the general binary expansion of $U_j$ and denote $\bU_D=(U_{1,D}, U_{2,D},\cdots, U_{p,D})^T$ as the depth-$D$ binary approximation of $\bU$. Let $\mathbf{B}^{p \times D}$ denote the set of all $p\times D$ binary matrices with entries being either $0$ or~$1$. We use a matrix $\Lambda=\Lambda^{p \times D} \in \mathbf{B}^{p \times D}$ to index an interaction of binary variables $\{A_{j,d}\}$ via  $A_{\Lambda}= \prod\nolimits_{j=1}^p\prod\nolimits_{d=1}^D (A_{j,d})^{\Lambda_{jd}}$. For the zero matrix $\Lambda=\zero^{p \times D},$ we define $A_{\zero^{p \times D}}=1.$ 

With the above notation, we develop the following theorem on the {\bf b}inary {\bf e}xpansion {\bf a}pproximation of {\bf u}niformi{\bf ty} (BEAUTY), which provides an approximation of the characteristic function of {\it any} distribution supported within $[-1,1]^p$ from the expectation of a polynomial of general binary expansion interactions.

\begin{theorem}[Binary Expansion Approximation of Uniformity, BEAUTY]\label{thm: beauty}
	Let $\bU$ be a $p$-dimensional random vector such that $U_j\in [-1,1], \forall j$. Let $\phi_{\bU}(\bt)$ be the characteristic function of $\bU$ for any $\bt=(t_1,\ldots,t_p)^T\in \RR^p$. We have 
	\begin{equation}\label{eq: bechar}
		e^{i \bt^T\bU_D}=\sum\nolimits_{\Lambda\in \mathbf{B}^{p \times D}} A_\Lambda\Psi_{\Lambda}(\bt) 
	\end{equation}
	and
	\begin{equation}\label{eq: beauty}
		\phi_{\bU}(\bt)=\Eb[\exp(i\bt^T\bU)] = \lim_{D\to\infty} \sum\nolimits_{\Lambda \in \mathbf{B}^{p \times D}} \Psi_{\Lambda}(\bt)\Eb[ A_{\Lambda}],
	\end{equation}
	where $\Psi_{\Lambda}(\bt) = \prod\nolimits_{j=1}^p\prod\nolimits_{d=1}^D \{\cos(t_j/2^{d})\}^{1-\Lambda_{jd}}\{i \sin(t_j/2^{d}) \}^{\Lambda_{jd}}.$
\end{theorem}  
 
Building upon Lemma~\ref{thm: bee}, \eqref{eq: bechar}  equates a complex exponent $e^{i \bt^T\bU_D}$ and a polynomial of binary variable $A_\Lambda$'s derived from the binary expansion of $\bU_D$. Equation~\eqref{eq: beauty}  reveals that the characteristic function of {\it any} random vector supported within $[-1,1]^p$ can be approximated by a linear combination of $\Psi_{\Lambda}(\bt)$'s, representing products of homogeneous trigonometric functions. As per Theorem 1.3 in \cite{zhao2023},  these functions are linearly independent. Furthermore, the coefficients of this linear combination correspond to the expectations of all binary variables in the $\sigma$-field induced by $\bU_D$. These expectations encapsulate the distributional properties of $\bU$, and   inference on them yields crucial insights into the distribution of $\bU$. Specifically, they provide clear insights about the goodness-of-fit tests by translating distributional properties into those of $\Eb[A_{\Lambda}]$'s. Moreover, sufficient statistics for $\Eb[A_{\Lambda}]$'s are the {\it symmetry statistics} $S_\Lambda=\sum_{i=1}^{n}A_{\Lambda,i}$ and equivalently $\bar{S}_\Lambda=n^{-1}\sum_{i=1}^{n}A_{\Lambda,i}.$ Consequently, test statistics should be constructed as a function of $\bar{S}_\Lambda$'s. We list some important examples connecting goodness-of-fit tests and $\Eb[A_{\Lambda}]$'s below:

\noindent {\it (1) Test of uniformity.} Consider the collection of non-zero $\Lambda$'s, $\cL_{p,D,\text{unif}}=\{ \Lambda \in \Bb^{p\times D}: \Lambda \neq \zero^{p \times D}\}$. Note that $\bU \sim \text{Unif}[-1,1]^p$ if and only if %$\Eb[A_\Lambda]=0$ for $\Lambda \in \lim_{D \rightarrow \infty}\cL_{p,D,\text{unif}},$ {\color{blue} This limit of a series of sets is not well defined. I suggest to change it as: 
$\Eb[A_\Lambda]=0$ for $\Lambda \in \cL_{p,D,\text{unif}}$ where $D$ is any positive integer. Consequently, %in which case 
\[\Eb[\exp(i\bt\bU)]= \lim_{D\to\infty} \prod\nolimits_{d=1}^{D} \Psi_{\zero^{p \times D}}(\bt)=\prod\nolimits_{j=1}^{p}\lim_{D\to\infty} \prod\nolimits_{d=1}^{D}  \{\cos(t_j/2^{d})\} = \prod\nolimits_{j=1}^{p} \{\sin(t_j)/t_j\},
\] 
i.e., \eqref{eq: beauty} recovers the characteristic function of \text{Unif}$[-1,1]^p.$ Therefore, the test of approximate uniformity \eqref{eq: test_ind_unif_D} is equivalent to test for any positive integer $D$,
\[ 
H_{0,D}: \Eb[A_\Lambda]=0,  \text{ for } \Lambda \in \cL_{p,D,\text{unif}}.
\]

\noindent {\it (2) Test of independence between two univariate random variables.}   The BEAUTY equation in Theorem \ref{thm: beauty} offers clear insights into the test of independence for bivariate copula up to a certain depth $D$. When $p_1=p_2=1$, it is always possible to transform $U_1$ and $U_2$ into uniform random variables on $[-1,1]$ using their marginal distributions. In this context, \cite{zhang2019bet} demonstrated that test \eqref{eq: test_ind} with $p_1=p_2=1$ considered the following test for any positive integer $D$,
$$
H_{0,D}: \Eb[A_{\Lambda}]=0 \text{ for } \Lambda \in \cL_{2,D,\text{cross}},
$$
where $\cL_{2,D,\text{cross}}=\{\Lambda=\Lambda_1 \rowbind \Lambda_2: \Lambda_1 \in \cL_{1,D,\text{unif}} \text{ and }\Lambda_2 \in \cL_{1,D,\text{unif}}\}$. Here, $\rowbind$ denotes the row binding of matrices with the same number of columns (see Definition \ref{def: rbind} in the supplement). 

\noindent {\it (3) Test of independence between response and predictors.}  The BEAUTY equation provides an in-depth understanding of testing the association between $p_1>1$ predictors $\bU_1$ and an arbitrary response $U_2$, extending beyond the scope of traditional regressions. As noticed before, $U_2$ can always be transformed into a uniform variable on $[-1,1]$. Note that $U_2$ is independent of $\bU_1$ if and only if $\sum_{\Lambda\in \bB^{(p_1+1)\times D}}\Eb[A_\Lambda] \Psi_{\Lambda}(\bt)=0$ for any positive integer $D$. Note also that from Theorem~1.3 in \cite{zhao2023}, the $\Psi_{\Lambda}(t)$'s are linearly independent in the function space. Therefore, by comparing the characteristic function of the joint distribution and the product of marginal characteristic functions, we see that for any positive integer $D$, the null hypothesis that $\bU_{1,D}$ and $U_{2,D}$ are independent is equivalent to the following hypothesis over the expectations of $A_\Lambda$'s:
:
$$
H_{0,D}: \Eb[A_{\Lambda}]=0  \text{ for } \Lambda \in \cL_{p_1+1,D,~\text{joint cross}},
$$ 
where $\cL_{p_1+1,D,~\text{joint cross}}=\{\Lambda=\Lambda_1 \rowbind \Lambda_2:  \Lambda_1 \in \cL_{p_1,D,\text{unif}} \text{ and }\Lambda_2 \in \cL_{1,D,\text{unif}}\}$.

%Thus, according to BEAUTY and the fundamental Theorem 1.3 from \cite{zhao2023} which states that the $\Psi_{\Lambda}(t)$'s are linearly independent in the function space, % \eqref{eq: bechar} indicates that 
%$U_2$ is independent of $\bU_1$ if and only if %$\lim_{D\to\infty}\sum_{\Lambda\in \bB^{(p_1+1)\times D}}\Eb[A_\Lambda] \Psi_{\Lambda}(\bt) = 0$ %{\color{blue} Again, I will suggest to replace this limit by:
%$\sum_{\Lambda\in \bB^{(p_1+1)\times D}}\Eb[A_\Lambda] \Psi_{\Lambda}(\bt)=0$ for any positive integer $D$.
%%{\color{blue}
%%In the three aforementioned examples, when the original hypotheses are true (i.e., (i) uniformity, (ii) bivariate independence, and (iii) independence between a univariate response and a vector), the corresponding hypotheses in terms of the binary expansion are also true. Furthermore, due to the linear independence of $\Psi_{\Lambda}(\mathbf{t})$'s, the converse is also true. In other words, these two sets of hypotheses are equivalent.
%%}

% {\color{red}[Wen: reviewer's question: is $\bU_1$ necessarily assumed to be uniform over $[-1,1]^{p_1}$?$\Downarrow$]
\noindent {\it (4) Test of independence between a uniform vector and an arbitrary vector.} For random vector $\bU_1\sim \mathrm{Unif}[-1,1]^{p_1}$ and arbitrary random vector $\bU_2$ within $[-1,1]^{p_2}$,  %respectively, if we know the joint distributions of one of them, then we can transform it to have uniform distribution of the hypercube. In this case, 
by leveraging Theorem~\ref{thm: beauty}, %to compare the characteristic function of the joint distribution with the product of the marginal characteristic distributions
the null hypothesis that posits the independence of $\bU_{1}$ and $\bU_{2}$ can be approximated by the following one over the expectations of $A_\Lambda$'s: 
$$
H_{0,D}: \Eb[A_{\Lambda}]=0  \text{ for } \Lambda \in \cL_{p_1+p_2,D,~\text{joint cross}},
$$ 
where $\cL_{p_1+p_2,D,~\text{joint cross}}=\{\Lambda=\Lambda_1 \rowbind \Lambda_2:  \Lambda_1 \in \cL_{p_1,D,\text{unif}} \text{ and }\Lambda_2 \in \cL_{p_2,D,\text{unif}}\}$.

\section{Unification of Several Tests of Independence}\label{sec: unification}
To construct a powerful test statistic, we first study existing tests of independence and their properties under the BET framework. We consider three important test statistics: Spearman's $\rho$ \citep{spearman1904proof}, the $\chi^2$ statistics, and the distance correlation \citep{szekely2007}. We find that each of these statistics can be approximated by a certain quadratic form of symmetry statistics. We further discuss the effect of the weight matrix in the quadratic forms on their power properties.

Since each specific statistic may involve a different collection of binary interactions, %to facilitate the description, 
we denote a collection of certain $\Lambda$'s by $\cL.$ For such a collection $\cL$, we denote the vector of $A_{\Lambda}$'s, $S_\Lambda$'s and $\bar{S}_\Lambda$'s with $\Lambda \in \cL$ by $A_\cL$, $\bS_\cL$ and $\bbS_\cL$, respectively. 

\subsection{Spearman's $\rho$}\label{subsec: spe}
As a robust version of the Pearson correlation, the Spearman's $\rho$ statistic  leads to a test with high asymptotic relative efficiency compared to the optimal test with Pearson correlation under bivariate normal distribution \citep{lehmann2006testing}. We show below it can be approximated by a quadratic form of symmetry statistics.

When $U_1$ and $U_2$ are marginally uniformly distributed over $[-1,1]$, Spearman's $\rho$ can be written as the correlation between $U_1$ and $U_2$, {\it i.e.}, 
\begin{equation}
	\rho=3\Eb[U_1U_2]=3\Eb\left[\sum_{d_1=1}^{\infty}\frac{A_{1,d_1}}{2^{d_1}}\sum_{d_2=1}^{\infty}\frac{A_{2,d_2}}{2^{d_2}}\right]=3\lim_{D \rightarrow \infty} \sum_{\Lambda \in \cL_{2,D,\text{spe}} } \br_D^T\Eb[A_{\Lambda}],
\end{equation}
where  $\cL_{2,D,\text{spe}}=\{\Lambda=\Lambda_1 \rowbind \Lambda_2: \Lambda_1, \Lambda_2 \in \Bb^{1\times D}, \text{ where } \Lambda_1\one=1 \text{ and } \Lambda_2\one=1\}$ consists of $2 \times D$ matrices whose rows are both binary vectors with only one unique $1$, and the $D^2$-dimensional vector $\br_D$ has entry $2^{-(d_1+d_2)}$ corresponding to $\Eb[A_{1,d_1}A_{2,d_2}].$ The test based on Spearman's $\rho$ rejects the null when the estimate of $\rho$ has a large absolute value. This test statistic can be approximated with 
$$Q_{\rho,D}=\frac{1}{n}(\br_D^T \bS_{\cL_{2,D,\text{spe}}})^2=\frac{1}{n}\bS_{\cL_{2,D,\text{spe}}}^T\br_D \br_D^T \bS_{\cL_{2,D,\text{spe}}},$$ 
which is a quadratic form with a rank-one weight matrix $\Wb_{\rho,D}=\br_D \br_D^T.$ 

Although the test based on Spearman's $\rho$ has a higher power against the linear form of dependency particularly present in bivariate normal distributions, we see from $\cL_{2,D,\text{spe}}$ and $\Wb_{\rho,D}$ that this test only considers $D^2$ out of $(2^D-1)^2$ cross interactions of binary variables in $\cL_{2,D,\text{cross}}$. Thus this test is not capable of detecting complex nonlinear forms of dependency.

\subsection{$\chi^2$ Test Statistic}\label{subsec: chi2}
When $U_1$ and $U_2$ are $\text{Unif}[-1,1]$ distributed, the binary expansion up to depth $D$ effectively leads to a discretization of $[-1,1]^2$ into a $2^D \times 2^D$ contingency table. Classical tests for contingency tables such as $\chi^2$-test can thus be applied. Similar tests include Fisher's exact test and its extensions \citep{ma2019fisher}. Multivariate extensions of these methods include \cite{gorsky2018multiscale,lee2023beret}.

In \cite{zhang2019bet}, it is shown that the $\chi^2$-statistic at depth $D$ can be written as the sum of squares of symmetry statistics for cross interactions. Thus,
$$Q_{\chi^2}=\frac{1}{n} \bS_{\cL_{2,D,\text{cross}}}^T \bS_{\cL_{2,D,\text{cross}}}$$
where $\cL_{2,D,\text{cross}}$ is the collection of all cross interactions. The weight matrix for $Q_{\chi^2}$ is thus the identity matrix $\Ib_{(2^D-1)\times (2^D-1)}$. 

The Max BET proposed in \cite{zhang2019bet} can be approximated by a quadratic form with another diagonal weight matrix, which we explain in the Supplementary Materials. These tests with diagonal weights can detect signals among the squared symmetry statistics, but might be powerless for signals from their cross products.

\subsection{Distance Correlation between uniformly distributed random vectors}\label{subsec: char}
To study the dependency between a $p_1$-dimensional vector $\bU_1$ and a $p_2$-dimensional vector $\bU_2$, in \cite{szekely2007},  a class of measures of dependence is defined as
\begin{equation}\label{eq: dcor}
	\cV^2(\bU_1,\bU_2)=\int_{\RR^{p_1+p_2}} |\phi_{(\bU_1,\bU_2)}(\bt_1,\bt_2)-\phi_{\bU_1}(\bt_1)\phi_{\bU_2}(\bt_2)|^2 w(\bt_1,\bt_2) d\bt_1d\bt_2,
\end{equation}
where $\phi_{(\bU_1,\bU_2)}(\bt_1,\bt_2)$ is the characteristic function of the joint distribution of $(\bU_1,\bU_2),$ $w(\bt_1,\bt_2)$ is a suitable weight function, and $\phi_{\bU_k}(\bt_k)$ is the characteristic function of $\bU_k, k=~1,2$. Note that $\cV^2(\bU_1,\bU_2)=0$ if and only if $\bU_1$ and $\bU_2$ are independent. The distance correlation is then defined through $\cV^2(\bU_1,\bU_2)$ and admits some desirable properties such as universal consistency against alternatives with finite expectation.

When $\bU_1 \sim \text{Unif}[-1,1]^{p_1}$ and $\bU_2 \sim \text{Unif}[-1,1]^{p_2}$, by Theorem~\ref{thm: beauty}, the term corresponding to $\Lambda=\zero$ cancels with $\phi_{\bU_1}(\bt_1)\phi_{\bU_2}(\bt_2)$, and we can write \eqref{eq: dcor} as
\begin{equation}
	\begin{split}
		\cV^2(\bU_1,\bU_2)=& \lim_{D \rightarrow \infty} \int_{\RR^{p_1+p_2}}\bigg|\sum_{\Lambda \in \cL_{p_1+p_2,D,\text{unif}}} \Psi_{\Lambda}(\bt)\Eb[ A_{\Lambda}]\bigg|^2 w(\bt_1,\bt_2) d\bt_1d\bt_2\\
		=& \lim_{D \rightarrow \infty} \sum_{\Lambda_1,\Lambda_2 \in \in \cL_{p_1+p_2,D,\text{unif}}} w_{\Lambda_1,\Lambda_2}\Eb[ A_{\Lambda_1}] \Eb[ A_{\Lambda_2}] \\
		=& \lim_{D \rightarrow \infty} \Eb[\bA_{\cL_{p_1+p_2,D,\text{unif}}}]^T\Wb_{\cV^2,p_1,p_2,D} \Eb[\bA_{\cL_{p_1+p_2,D,\text{unif}}}] 
	\end{split}
\end{equation}
where $\cL_{p_1+p_2,D,\text{unif}}=\{\Lambda \in \Bb^{(p_1+p_2)\times D}: \Lambda \neq \zero^{p \times D}\}$, and the weight matrix $\Wb_{\cV^2,p_1,p_2,D}$ consists of constants $w_{\Lambda_1,\Lambda_2}$'s from the integration over $\bt_1$ and $\bt_2$. The test is significant when the empirical quadratic form $Q_{\cV^2,p1,p2,D}$ is large, where
$$Q_{\cV^2,p1,p2,D}=\frac{1}{n} \bS_{\cL_{p_1+p_2,D,\text{unif}}}^T\Wb_{\cV^2,p_1,p_2,D} \bS_{\cL_{p_1+p_2,D,\text{unif}}}.$$  

Note that $\Wb_{\cV^2,p_1,p_2,D}$ here depends only on the weight function $w(\bt_1,\bt_2)$ and is deterministic. Hence, the test based on $Q_{\cV^2,p1,p2,D}$ will have a high power when the vector of $\Eb[A_{\Lambda}]$'s from the alternative distribution lies in the subspace spanned by eigenvectors of $\Wb_{\cV^2,p_1,p_2,D}$ corresponding to its largest eigenvalues. On the other hand, if instead the signals lie in the subspace spanned by eigenvectors of $\Wb_{\cV^2,p_1,p_2,D}$ corresponding to its lowest eigenvalues, then the power of the test could be considerably compromised. Therefore, a deterministic weight over symmetry statistics becomes a general uniformity issue of existing test statistics. In the next section, we study data-adaptive weights with the aim to improve the power by setting proper weights both among diagonal and off-diagonal entries in the matrix.

\section{The BEAST and Its Properties}\label{sec: beast}

\subsection{The First Two Moments of Binary Interactions}\label{subsec: mv}
The unification in Section~\ref{sec: unification} inspires us to consider a class of nonparametric statistics for the goodness-of-fit test as a weighted sum of symmetry statistics. Since the properties of this form of statistics are closely related to the first two moments of the binary interaction variables in the filtration, we consider the collection of all nontrivial binary interactions $\cL=\cL_{p,D,\text{unif}}=\{ \Lambda \in \Bb^{p\times D}: \Lambda \neq \zero^{p \times D}\}$ and study the moment properties of the corresponding binary random vector $\bA_\cL.$

We begin by studying the connection between the $(2^{pD}-1)\times 1$ vector $\bA_\cL$ and the multinomial distribution from the corresponding discretization with $2^{pD}$ categories. We order the indices $\Lambda$'s in $\bA_\cL$ by the integer corresponding to the binary vector representation $vec(\Lambda^T),$ where $vec(\cdot)$ is the vectorization function. For example, the last ({\it i.e.} the $(2^{pD}-1)$th) entry in $\bA_\cL$ corresponds to the $\Lambda=\one^{p\times D}$. We also denote the $2^{pD}\times 1$ vector of cell probabilities in the multinomial distribution by $\bp_c.$  Label the entries in $\bp_c$ by binary matrices $\Lambda \in \Bb^{p \times D}$ through $\Lambda=\Lambda_1 \rowbind \ldots \rowbind \Lambda_p,$ where each realization of the $2^D \times 1$ vector $\Lambda_j$ labels one of the $2^D$ intervals for dimension $j$ from low to high according to 1 plus the integer corresponding to the binary representation of $\Lambda_j^T$. We define a $2^{pD}\times 1$ random vector $\bZ=(Z_\Lambda) \sim Multinomial (1,\bp_c)$ to denote one draw from the $2^{pD}$ intervals from the discretization. With the above notation, we develop the general binary interaction design (BID) equation, which extends the two-dimensional case in \cite{zhang2019bet}.

\begin{theorem}\label{thm: gbid}
	Let $\bA_c=(1,\bA_\cL^T)^T,$ $\bmu_{c}=\Eb[\bA_c]$ and $\bSigma_{\bmu_{c}}=\Eb[\bA_c\bA_c^T].$ Denote the $2^{pD}\times 2^{pD}$ Sylvester's Hadamard matrix by $\Hb.$ We have the binary interaction design (BID) equation
	\begin{equation}\label{eq: gbid}
		\bA_c=\Hb \bZ.
	\end{equation}
	In particular, we have the BID equation for the mean vector
	\begin{equation}\label{eq: gbid_mu}
		\bmu_{c}=\Hb\bp_c
	\end{equation}
	and the corresponding BID equation for $\bSigma_{\bmu_{c}}$
	\begin{equation}\label{eq: gbid_sig}
		\bSigma_{\bmu_{c}}=\Hb diag(\bp_c) \Hb,
	\end{equation}
	where $diag(\bp_c)$ is the diagonal matrix with diagonal entries corresponding to $\bp_c$.
\end{theorem}

The Hadamard matrix $\Hb$ is also referred to as the Walsh matrix in engineering, where the linear transformation with $\Hb$ is referred to as the Hadamard transform \citep{lynn1973introduction,golubov2012walsh,harmuth2013transmission}. The earliest referral to the Hadamard matrix we found in the statistical literature is \cite{Pearl1971}, and it is also closely related to the orthogonal full factorial design \citep{cox2000theory,box2005statistics}. In our context of testing independence, the BID equation can be regarded as a transformation from the physical domain to the frequency domain, which turns the focus to global forms of non-uniformity instead of local ones. In developing statistics, this transformation facilitates regularizations through thresholding, as $\bmu_{\cL}=\zero$ is equivalent to uniformity $\bp_c=1/2^{pD}\one.$ This transformation also enables clear interpretations of statistical significance with the form of dependency, as shown in \cite{zhang2019bet}.

To study the power of the test of uniformity, we further study the properties of the first two moments of $\bA_\cL.$ Let $\bmu_{\cL}=\Eb[\bA_\cL]$ and $\bSigma_{\bmu_{\cL}}=\Eb[\bA_\cL\bA_\cL^T]$ denote the vector of expectations and the matrix of second moments of $\bA_\cL$ respectively.  We summarize some properties of $\bmu_{\cL}$ and $\bSigma_{\bmu_{\cL}}$ in the following theorem.

\begin{theorem}\label{thm: musigma}
	We have the following results on the properties of first two moments of binary interaction variables in the binary expansion filtration.
	\begin{enumerate}[(a)]
		\item The connection between the first and second moments of binary interactions:
		\begin{equation}\label{eq: bem1m2}
			\bmu_{c}^T\bSigma_{\bmu_{c}}^{-1}\bmu_{c}=1.
		\end{equation} 
		\item The connection between the harmonic mean of probabilities and the Hotelling's $T^2$ quadratic form when $p_{\Lambda}>0,\forall \Lambda \in \cL$:
		\begin{equation}\label{eq: harmonichotelling}
			\frac{1}{2^{2pD}}\sum_{\Lambda \in \cL} p_{\Lambda}^{-1}=1+\bmu_{\cL}^T(\bSigma_{\bmu_{\cL}}-\bmu_{\cL}\bmu_{\cL}^T)^{-1}\bmu_{\cL}=(1-\bmu_{\cL}^T\bSigma_{\bmu_{\cL}}^{-1}\bmu_{\cL})^{-1}.
		\end{equation}
		\item For $\bmu_{\cL}$ with $\|\bmu_{\cL}\|_2 \le (2^{pD}-1)^{-1/2},$ with constant $c_{p,D}=(2^{pD}-2)/\sqrt{2^{pD}-1},$
		\begin{equation}\label{eq: 3qf}
			\|\bmu_{\cL}\|_2^2-c_{p,D}\|\bmu_{\cL}\|_2^3 \le \bmu_{\cL}^T \bSigma_{\bmu_{\cL}} \bmu_{\cL} \le \|\bmu_{\cL}\|_2^2+c_{p,D}\|\bmu_{\cL}\|_2^3.
		\end{equation} 
		\item Denote the vector-valued function $(\bSigma_{\bmu_{\cL}}-\bmu_{\cL}\bmu_{\cL}^T)^{-1}\bmu_{\cL}$ by $\bg(\bmu_{\cL})=(g_\Lambda(\bmu_{\cL}))$ for each $\Lambda \in \cL$. As $\norm{\bmu_{\cL}}_2\rightarrow 0,$ 
		\begin{equation}\label{eq: muasymp}
			g_\Lambda(\bmu_{\cL})=\mu_{\Lambda}+ o(\norm{\bmu_{\cL}}_2).
		\end{equation}
	\end{enumerate}
\end{theorem}
To the best of our knowledge, the results in Theorem~\ref{thm: musigma}, despite their simplicity, have not been documented in literature. These simple results unveil interesting insights of the first two moments of binary variables in the filtration. The quadratic form in \eqref{eq: bem1m2} characterizes the functional relationship between $\bmu_{c}$ and $\bSigma_{\bmu_{c}}$. The two equations in \eqref{eq: harmonichotelling} show that for binary variables, the Hotelling $T^2$ quadratic form is a monotone function of the harmonic mean of the cell probabilities in the corresponding multinomial distribution. The inequalities in \eqref{eq: 3qf} reveal the eigen structure of $\bSigma_{\bmu_{\cL}}$ when the signal $\bmu_{\cL}$ is weak. The Taylor expansion in \eqref{eq: muasymp} provides the asymptotic behavior of $(\bSigma_{\bmu_{\cL}}-\bmu_{\cL}\bmu_{\cL}^T)^{-1}\bmu_{\cL}$ when the joint distribution is close to the uniform distribution. These insights shed important lights on how we can develop a powerful goodness-of-fit test, as we explain in Sections~\ref{subsec: oracle} and \ref{subsec: beast}.

%({\color{red} $\uparrow$ add theorem on empirical cdf, using BET Theorem 4.2})

\subsection{An Oracle Approach for Test Construction}\label{subsec: oracle}
In this section, we study how to construct a powerful robust nonparametric test of uniformity based on what we learned in Sections~\ref{sec: unification} and~\ref{subsec: mv}. As discussed in Section~\ref{sec: unification}, the deterministic weights of symmetry statistics in existing tests create an issue on the uniformity and robustness: They make the test powerful for some alternatives but not for others. Therefore, we construct a test statistic with data-adaptive weights, which allow the test to adjust itself towards the alternative to improve the power. We refer this class of statistics as the {\bf b}inary {\bf e}xpansion {\bf a}daptive {\bf s}ymmetry {\bf t}est (BEAST).

We construct our test through an oracle approach. Suppose we know from an oracle $\bmu_{\cL}$ and thus $\bSigma_{\bmu_{\cL}}$ as shown in Theorem \ref{thm: gbid}. Note that by Theorem~4.1, 4.2 and 4.3 in \cite{zhang2019bet}, under $H_{0,D}$ of \eqref{eq: test_ind_unif_D}, the distribution of $\bbS_{\cL}$ is either pairwise independent centered binomial or uncorrelated centered hypergeometric, depending on whether the marginal distributions are known. Therefore, for fixed $p$ and $D$, with a large $n$ and the central limit theorem on $\bbS_\cL=\bS_\cL/n$, we approximately have a simple-versus-simple hypothesis testing problem:
$$H_0: \sqrt{n}\bbS_\cL \sim \cN(\zero,\Ib) \text{ v.s. } H_1: \sqrt{n}(\bbS_\cL-\bmu_{\cL}) \sim \cN(\zero,\bSigma_{\bmu_{\cL}}-\bmu_{\cL}\bmu_{\cL}^T).$$
According to the fundamental Neyman-Pearson Lemma \citep{neyman1933ix}, the corresponding most powerful (MP) test is the likelihood ratio test. We thus consider the data-relevant part of the log-likelihood ratio of the above two distributions,
$$f_{\bbS_\cL}(\bmu_{\cL})=-\frac{1}{2n}\bS_\cL^T(\Ib-(\bSigma_{\bmu_{\cL}}-\bmu_{\cL}\bmu_{\cL}^T)^{-1})\bS_\cL+\bmu_{\cL}^T(\bSigma_{\bmu_{\cL}}-\bmu_{\cL}\bmu_{\cL}^T)^{-1}\bS_\cL.$$
For a large $n$, the dominating term in $f_{\bbS_\cL}(\bmu_{\cL})$ is $\bmu_{\cL}^T(\bSigma_{\bmu_{\cL}}-\bmu_{\cL}\bmu_{\cL}^T)^{-1}\bS_\cL.$ By \eqref{eq: muasymp} in Theorem~\ref{thm: musigma}, the first order Taylor expansion of this term is precisely $\bmu_{\cL}^T\bS_\cL$! This implies that the MP test rejects when $\bbS_\cL$ is colinear with $\bmu_{\cL}.$ The above heuristics thus suggests that we consider the oracle test statistic $B_{\text{oracle}}=\bmu_{\cL}^T\bbS_\cL/\|\bmu_{\cL}\|_2$.

In our simulation studies in Section~\ref{sec: simu}, since we know the form of the alternative distribution, we can estimate $\bmu_{\cL}$ with high accuracy through an independent simulation. That is, with the known alternative distribution $\Pb_{\bU}$, for a large $K$ we simulate $\bV_1,\ldots,\bV_K \stackrel{i.i.d.}{\sim}\Pb_{\bU}.$ From the binary expansion of $\bV_1,\ldots,\bV_K,$ we obtain the vector of symmetry statistics $\Tilde{\bS}_{\cL}$ and an estimate of $\bmu_{\cL}$ denoted by $\Tilde{\bmu}_{\cL}=\Tilde{\bS}_{\cL}/n.$ The oracle test statistic from simulations is then $\tilde{B}_{\text{oracle}}=\Tilde{\bmu}_{\cL}^T \bbS_{\cL}/\|\Tilde{\bmu}_{\cL}\|.$

We show in simulations that even when $D$ is as small as $3$, $\tilde{B}_{\text{oracle}}$ is  powerful, and numerically it outperforms all existing competitors under consideration across a wide spectrum of alternatives and noise levels. For example, for the cases when the joint distributions are Gaussian with linear dependency, the power curves of $\tilde{B}_{\text{oracle}}$ dominate those of the distance correlation when $p=2$ and the F-test when $p=3,$ which are known to be optimal. Compared to existing tests, the gain of the BEAST with oracle in power suggests that suitably chosen deterministic weights for the alternative provide a unified yet simple solution to improve the power. To the best of our knowledge, this is the first time that such a benchmark on the feasible power performance is available for the problem of testing uniformity.

Besides the useful insight about the feasible limit of power, the oracle also provides insights on the optimal weights under each alternative. For example, in simulations we find high colinearity between the approximate oracle weight vector $\Tilde{\bmu}_{\cL}$ and that of the Spearman's $\rho$, $\br_D$, as found in Section~\ref{subsec: spe}. This weight vector makes the one-sided test with $\tilde{B}_{\text{oracle}}$ more powerful than the two-sided test with Spearman's $\rho.$

Although the optimal weight $\bmu_\cL$ or $\Tilde{\bmu}_{\cL}$ is unknown in practice, an unbiased and asymptotically efficient estimate of $\bmu_\cL$ is $\bbS_\cL.$ This motivates us to develop an approximation of $\tilde{B}_{\text{oracle}}$ through resampling and regularization, which we discuss next.

\subsection{The BEAST Statistic}\label{subsec: beast}
In practice, we are agnostic about $\bmu_{\cL}.$ Blindly replacing $\Tilde{\bmu}_\cL$ in $\tilde{B}_{\text{oracle}}$ with $\bbS_\cL$ will result in colinearity with itself and the statistic reduces to the classical $\chi^2$-test statistic. Traditionally, the data-splitting strategy has often been employed for this type of situations to facilitate data-driven decision \citep{Hartigan1969,Cox1975}, {\it i.e.}, half of the data is used to calibrate the statistical procedure such as screening the null features \citep{WassermanandRoeder2009,BarberandCandes2019}, determining the proper weights for individual hypotheses \citep{Ignatiadis2016}, recovering the optimal projection for dimension reducion \citep{Huang2015}, and estimating the latent loading for factor models \citep{FARM2019}, while a statistical decision is implemented using the remaining half. However, the single data-splitting procedure only uses half of the data for decision making, which inevitably bears undesirable randomness and therefore leads to power loss for hypothesis testing. Some recent efforts have shown that this shortcoming can be lessened by using multiple splittings \citep{RomanoDiCiccio2019,LiuYuLi2020,DaiLinXingLiu2020}. 

Motivated by the principle of multiple splitting, we propose to approximate $B_{\text{oracle}}$ through resampling: We replace $\bmu_{\cL}$ in $B_{\text{oracle}}$ with $\bbS_{\cL}$, and we replace $\bbS_{\cL}$ in $B_{\text{oracle}}$ with its resampling version $\bbS_{\cL}^*$. Important resampling methods include bootstrap \citep{efron1994introduction} and subsampling  \citep{Politis.etal1999}. Bootstrap and subampling are known to have similar performance in approximating the sampling distribution of the target statistic. In this paper, we use the subsampling method to facilitate the calculation of the empirical copula distribution when the marginal distributions are unknown \cite{nelsen2007introduction}. In addition to the above consideration, one intuition behind this resampling approach is to help distinguish the alternative distribution from the null: Under the null, since $\bmu_{\cL}=\zero,$ we expect the magnitude of $\bbS_{\cL}$ and $\bbS_{\cL}^*$ to be small and not very colinear after regularization. On the other hand, under the alternative, since $\bmu_{\cL}\neq \zero,$ we expect the two estimations of $\bmu_{\cL}$ to be both colinear with $\bmu_{\cL}$ and thus to be highly colinear themselves. Therefore, the magnitude of the test statistic could be different to help distinguish the alternative distribution from the null.

In addition, we apply regularization to accommodate sparsity, {\it i.e.}, the non-uniformity can be explained by a few binary interactions $\Lambda$'s with $\Eb[A_{\Lambda}]\neq 0.$ This sparsity assumption is often reasonable in the BET framework, since $\Eb[A_{\Lambda}]= 0$ is equivalent to the symmetry of distribution according to the interaction $\Lambda$. Thus, sparsity over $\Eb[A_{\Lambda}]$'s is equivalent to a highly symmetric distribution. For example, if a multivariate distribution is symmetric in every direction, then each one-dimensional projection of this distribution has a real characteristic function. By Theorem~\ref{thm: beauty}, we have $\Eb[A_{\Lambda}]=0$ for all $\Lambda$ involving an even number of binary variables ($\one_p^T\Lambda\one_D$ is even). Many global forms of dependency also correspond to sparse structures in $\bmu_{\cL}.$

The estimation of $\bmu_{\cL}$ under the sparsity assumption is closely related to the normal mean problem, where many good regularization based methods are readily available \citep{wasserman2006all}. For example, in \cite{donoho1994ideal}, it is shown that estimation with soft thresholding is nearly optimal. We denote the vector-valued soft thresholding function by $\cT(\bx,\lambda)$ for $q\times 1$ vector $\bx$ and threshold $\lambda>0,$ so that $\cT_\ell(\bx,\lambda)=\text{sign}(x_\ell)(|x_\ell|-\lambda)_+, ~\ell =1,\ldots,q.$ In construction of our test statistic, we choose to use soft-thresholding as a regularization step to screen the small observations in $\bbS_{\cL}$ and $\bbS_{\cL}^*$ due to the null distribution or due to the sparsity $\Eb[A_{\Lambda}]=0$ for certain interaction $\Lambda$'s under the alternative, thus improves the power of the test statistic.

In summary, we consider the approximation of $B_{\text{oracle}}$ through subsampling, while using regularization to obtain a good estimate of the optimal weight vector $\cT(\bbS_\cL,\lambda)/\|\cT(\bbS_\cL,\lambda)\|_2.$ The detailed steps are listed below.
\begin{enumerate}
	\item[Step 1:] From $n$ observations of $\bU_1,,\ldots,\bU_n,$ obtain $m$ subsamples of size $r$: $\bU_{1,k}^*,\ldots,\bU_{r,k}^*$, $k=1,\ldots,m.$ For each subsample $k,$ base on the binary expansions of $\bU_{1,k}^*,\ldots,\bU_{r,k}^*$, find the vector of average symmetry statistics $\bbS_{\cL,k}^*$. \item[Step 2:]Take the average over $m$ subsamples to obtain $\bbS_\cL^*={m}^{-1} \sum_{k=1}^{m} \bbS_{\cL,k}^*.$ Apply the soft-thresholding function to get an estimate of $\bmu_{\cL}$ as $\cT(\bbS_\cL^*,\lambda).$
	\item[Step 3:] The BEAST statistic $B_{\lambda}$ is obtained as
	\begin{equation}
		B_{\lambda}=\cT(\bbS_\cL,\lambda)^T\cT(\bbS^*_\cL,\lambda)/\|\cT(\bbS_\cL,\lambda)\|_2.
	\end{equation}
\end{enumerate} 
We study the empirical power of the BEAST in Section~\ref{sec: simu}, which shows that by approximating $\tilde{B}_{\text{oracle}}$ with regularization and subsampling, $B_{\lambda}$ has a robust power against many alternative distributions, especially complex nonlinear forms of dependency. 

We now study the asymptotic distributional properties of $B_{\lambda}$ under the assumption of known marginal distributions. Denote the $2^{pD}\times 1$ vector of cell proportions of the discretization out of $n$ samples by $\hat{\bp}_c.$ We have the following theorem on the distribution of the subsample symmetry statistic $\bbS_{\cL}$ condition on $\bbS_{\cL}$.

\begin{theorem}\label{thm: bbs}
	Condition on $\bbS_{\cL},$ as $m\rightarrow \infty,$ we have
	$$\sqrt{m}(\bbS_{\cL}^*-\bbS_{\cL}) \sim \cN\bigg(\zero, \frac{n-r}{r(n-1)}(\Hb diag(\hat{\bp}_c) \Hb-\bbS_{\cL}\bbS_{\cL}^T)_{[-1,-1]}\bigg)$$
	where $M_{[-1,-1]}$ is the submatrix of $M$ with the first row and first column removed. 
\end{theorem}

Theorem~\ref{thm: bbs} holds both under the null distribution and the alternative distribution. This result thus provides useful guidance and efficient algorithms to simulate the null and alternative distributions of $B_{\lambda}$ for any $\lambda.$ The detailed asymptotic distribution of $B_{\lambda}$ with a positive $\lambda$ and the analysis of the power function are useful for developing optimal adaptive tests and are interesting problems for future studies.

\subsection{Practical Considerations}
In this section, we discuss some practical considerations in applying the BEAST. The first practical issue is whether using the empirical CDF would lead to some loss of power. As discussed in \cite{zhang2019bet}, the difference between using the known CDF and empirical CDF is similar to the difference between the multinomial model and the multivariate hypergeometric model for the contingency table, in which the theory and performance are similar too. In all of our numerical studies, we considered the method using the empirical CDF.

A related issue is the choice of depth $D$ and threshold $\lambda$ in practice. In our simulations, we find that with $D=3$, the BEAST with oracle has a higher power than the linear model based tests for Gaussian data, which indicates that $D=3$ is sufficiently large to detect many important forms of dependency. In our simulation studies in Section 5 below and Section D in the supplmentary materials, we also show the power is relatively stable as $D$ increases slightly. Moreover, data studies show that using $D=2$ can already provide many interesting findings \cite{xiang2022pairwise} in practice. Therefore, we choose $D=3$ for this paper. We shall also choose a $\lambda=O(\sqrt{Dp/n})$ according to the large deviation of the symmetry statistic. That is approximately $90\%$ of these symmetry statistics under the null of uniformity will be thresholded to 0. A general optimal choice of $D$ and $\lambda$ for some specific alternative should come from a trade-off between them and $n,p,$ and the signal strength. This would be an interesting problem for future studies.

%=================================================================================================%

\section{Simulation Studies}\label{sec: simu}
\subsection{Testing Bivariate Independence}\label{subsec: p2}
In this section, we consider the problem of testing the bivariate independence. The sample size is set to be $n=128$. The BEAST with oracle and the BEAST are constructed with the empirical copula distribution and with $\cL_{2,D,\text{cross}}=\{\Lambda=\Lambda_1 \rowbind \Lambda_2:  \Lambda_1 \in \cL_{1,D,\text{unif}} \text{ and }\Lambda_2 \in \cL_{1,D,\text{unif}}\}, m=128$, $D=3$, $r=24$, and $\lambda=\sqrt{(pD\log 2)(8n)^{-1}}=0.064$. For the BEAST with oracle, we choose $K=10^5$ to obtain the oracle weights $\tilde{\bmu}_{\cL}$ and $\tilde{B}_{\text{oracle}}$ for each alternative distribution, and the critical region is then obtained through $10^4$ draws from the bivariate uniform distribution over $[0,1]^2$. For the BEAST, {the critical region is formed with $100$ $B_{\lambda}$'s permuted from the data}. The level of all tests is set to be $0.1.$

We compare the power of the two versions of the BEAST with the following methods: the $\chi^2$-test and its improvement the $U$-statistic permutation (USP) test \citep{berrett2021optimal, berrett2021usp} with the same discretization for $B_{\lambda}$, the Fisher exact scanning \citep{ma2019fisher}, the distance correlation \citep{szekely2007}, the $k$-nearest neighbor mutual information (KNN-MI, \cite{kinney2014equitability}) with the default parameters, the $k$-nearest neighbour based Mutual Information Test (MINT, \cite{berrett2019nonparametric}) with default averaging over $k$, MaxBET \citep{zhang2019bet}, BERET \citep{lee2023beret}, and the high-dimensional multinomial test (HDMultinomial) by \cite{Balakrishnan2019hypothesis}.  Among these tests, the HDMultinomial, the MINT, and the USP test are shown to be minimax optimal in power.

\begin{figure}[h!]
	\begin{center}
		\includegraphics[width=0.975\textwidth,height=9.25cm]{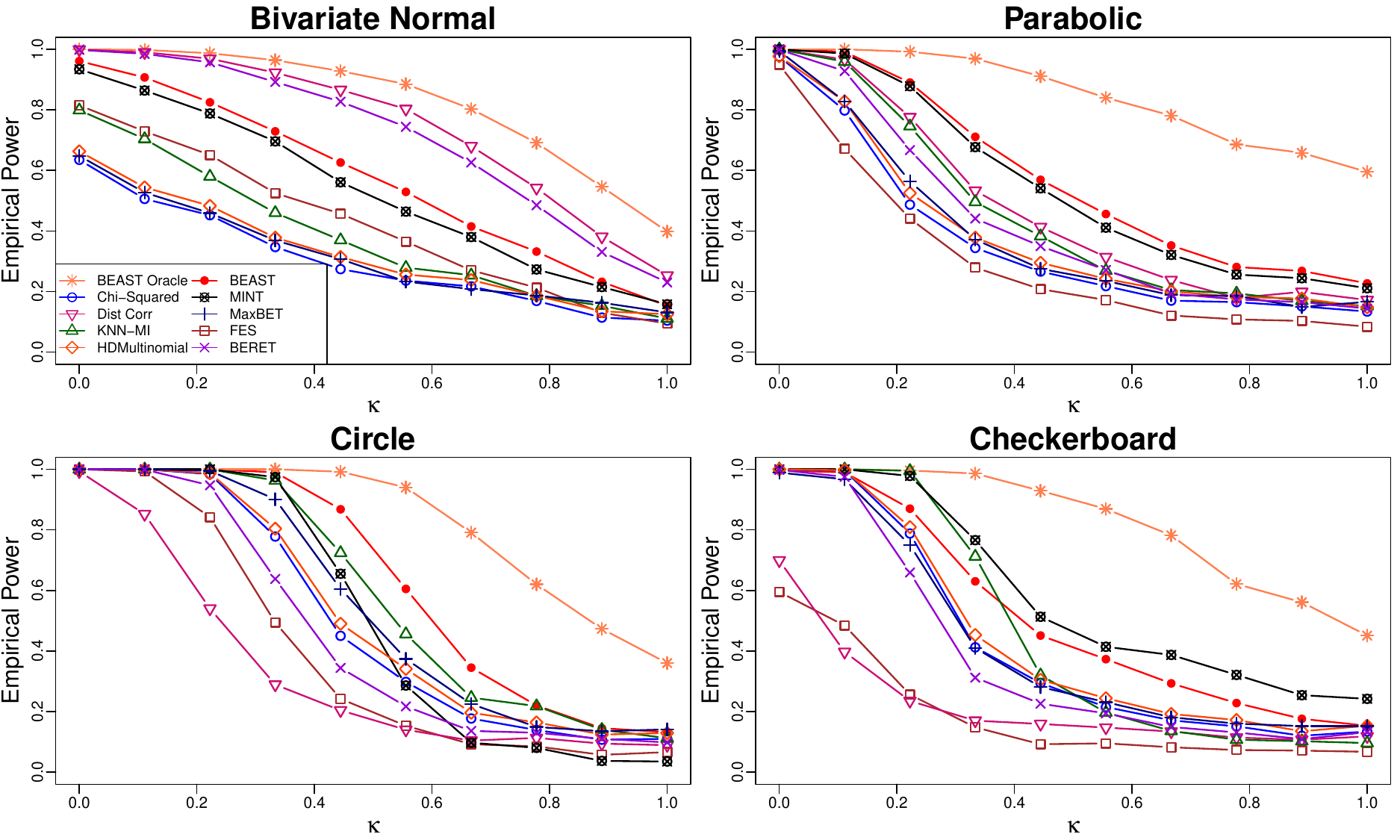}
	\end{center}
	\caption{The power curves of various methods when testing the bivariate independence under four alternatives. The sample size $n=128$ and the depth of the BEAST is chosen as $3$. The level of significance is set to be $0.1$. The BEAST with oracle provides a benchmark on the feasible power for all cases. The power of the BEAST consistently ranks within the top three among all tests for all cases, while being the best under the ``Parabolic'' and ``Circle'' cases. }\label{fig: p2d3}
\end{figure}

The data $(x_i, y_i), i=1,2,\cdots, n=128$ of the alternative distributions are generated according to the four different settings in Table~\ref{table: p2power} below. Parameter $\kappa$ is evenly spaced over $[0,1]$ to represent the level of noise. The settings are chosen such that the power curves display a thorough comparison for different signal strengths. In Figure~\ref{fig: p2d3}, $1,000$ simulations are conducted to calculate the empirical power of each test for each setting with a given $\kappa$.

\begin{table}
	\label{table: p2power} 
	%\resizebox{\textwidth}{!}{
	\caption{Simulation scenarios for $p=2$: The following variables are all independent. $\epsilon_j \sim \cN(0,1)$ for $j=1,\ldots,8$; $U \sim \text{Unif}[-1,1]$ ; $\vartheta \sim \text{Unif}[-\pi,\pi]$; $W \sim \text{Multi-Bern}(\{1,2,3\}, (1/3,1/3,1/3))$; $V_1 \sim \text{Bern}(\{2,4\},(1/2,1/2))$; and $V_2 \sim \text{Multi-Bern}(\{1,3,5\}, (1/3,1/3,1/3))$. $\kappa$ is evenly spaced between 0 and 1.}
	%}
	\def\arraystretch{1}%
	\centering
	\resizebox{\textwidth}{!}
	{\footnotesize 
		\begin{tabular}{l|l|l} 
			\toprule Scenario & Generation of $X$ & Generation of $Y$\\ \midrule
			Bivariate Normal  &  \multirow{ 1}{*}{$\displaystyle X=\sqrt{0.4-0.3\kappa}\epsilon_1+\sqrt{0.6+0.3\kappa}\epsilon_2$} & \multirow{ 1}{*}{$ \displaystyle Y=\sqrt{0.4-0.3\kappa}\epsilon_1+\sqrt{0.6+0.3}\epsilon_3 $} \\[0.5ex]
			Parabolic & $X=U~~~~~$ & $  Y=0.25X^2+(0.4\kappa+0.1)\epsilon_4$\\[0.5ex]
			Circle & $X=\cos \vartheta+(0.6\kappa+0.1)\epsilon_5~~~~~$ & $  Y=\sin\vartheta+(0.6\kappa+0.1)\epsilon_6$\\[0.5ex]			
			Checkerboard & $X=W+(0.3\kappa+0.05)\epsilon_{7}~~~~~$ & $ \displaystyle Y=
			V_1\mathbb{I}(W=2)+V_2\mathbb{I}(W\neq 2)+(1.2\kappa+0.2)\epsilon_{8}$\\
			\bottomrule
		\end{tabular}
	} 		 
\end{table}

We first comment on the performance of the BEAST with oracle. Although this test is not achievable in practice, it provides many important insights in these simulation examples. From Figure~\ref{fig: p2d3}, we see that with a small depth $D=3$, the BEAST with oracle achieves the highest power among all methods, for every alternative distribution and every level of noise. In particular, under the bivariate normal case, the power curve of $\tilde{B}_{\text{oracle}}$ is higher than that of the distance correlation, while leaving substantial gaps to other nonparametric tests. The good performance of the distance correlation is expected, since it has been shown that it is a monotone function of Pearson correlation under normality \citep{szekely2007}. These facts thus again show that the BEAST with oracle can accurately approximate the optimal power under an alternative. Therefore, the BEAST with oracle provides a useful benchmark for the performance of tests. 

Moreover, in this case we find high colinearity between the approximate oracle weight vector $\Tilde{\bmu}_{\cL}$ and that of the Spearman's $\rho$, $\br_D$, as found in Section~\ref{subsec: spe}. This shows the ability of $\tilde{B}_{\text{oracle}}$ to approximate the optimal weights. The higher power of $\tilde{B}_{\text{oracle}}$ can be also attributed to knowing the sign of correlation under this oracle.

The optimality of the BEAST with oracle is further demonstrated in other three more complicated scenarios with nonlinear dependency, where its power curve dominates all others by a huge margin. This result again indicates the potential of gains in power for these alternatives. To the extent of our knowledge, the BEAST with oracle is the first method in literature that evidences the potential of profound improvement in power via a suitable choice of weights.

We now turn to the comparison of $B_{\lambda}$ with existing tests. The general phenomenon in Figure~\ref{fig: p2d3} is that every existing test has some advantageous and disadvantageous scenarios. For examples, the Spearman's $\rho$ will have optimal power under the ``Bivariate Normal'' case while being powerless in the other three situations due to a zero correlation, the $\chi^2$-test has a good power in the ``Checkerboard'' scenario but has the worst power under the ``Bivariate Normal'' case, and the distance correlation has a high power under the ``Bivariate Normal'' and ``Parabolic'' cases while not performing well in the other two. These phenomena about the power properties of these three tests can be explained by the deterministic weight matrices in the approximate quadratic form of symmetry statistics, as discussed in Section~\ref{sec: unification}.

The empirical power of the BEAST, however, is always high against each alternative distribution and consistently ranks within the top three among all tests, for all alternatives, and for all levels of noise. In particular, the power curve of $B_{\lambda}$ dominates those of other tests under the scenarios `Parabolic'' and ``Circle.'' The reasons for this high power include (a) the subsampling approximation of the optimal weights $\bmu_{\cL}$ and the approximate MP test statistic $\tilde{B}_{\text{oracle}}$ and (b) the regularization step with soft-thresholding which takes advantage of the equivalence of sparsity and symmetry.

Note also that under the ``Checkerboard'' scenario, the data contain several natural clusters. This feature of the alternative distribution would favor statistical methods from the $k$-nearest neighbour methods. Therefore, the good powers of KNN-MI and MINT are expected. The fact that $B_{\lambda}$ has competitive power with KNN-MI and MINT under this scenario again demonstrates the ability of the BEAST to provide a high power despite being agnostic of the specific alternative. 

\subsection{Testing Independence of Response and Predictors}\label{subsec: p3}
In this section, we consider the test of independence between a bivariate predictor $(X_1,X_2)$ and one univariate response $Y$. Based on the BEAUTY equation in Theorem~\ref{thm: beauty} and the discussions afterwards, this test at depth $D$ is equivalent to test $H_0: \Eb[A_{\Lambda}]=\zero$ for 
$\Lambda \in \cL_{3,D,\text{cross}}$ where 
\begin{equation}\label{eqn:l3:cross}
	\cL_{3,D, ~\text{joint cross}} =\left\{ \Lambda=\Lambda_1 \rowbind \Lambda_2: \Lambda_1 \in \cL_{2,D, \text{unif}}, \Lambda_2 \in \cL_{1,D,\text{unif}} \right\}.
\end{equation}
Thus, $\tilde{B}_{\text{oracle}}$ and $B_{\lambda}$ are constructed according to $\cL_{3,D,\text{cross}}.$ The critical regions of these statistics are obtained through {permutations} similarly to that in Section~\ref{subsec: p2}. With $D=3$ and $p=3$, we set $\lambda=\sqrt{(pD\log 2)(8n)^{-1}}=0.078$ for the BEAST.

We compare $\tilde{B}_{\text{oracle}}$ and $B_{\lambda}$ with existing nonparametric tests of independence for vectors including the $\chi^2$-test from the same discretization for $B_{\lambda}$ with simulated $p$-values, the $F$-test from the linear model of $Y$ against $(X_1,X_2)$, the distance correlation \citep{szekely2007}, the $k$-nearest neighbor mutual information (KNN-MI, \cite{kinney2014equitability}) with the default parameters, the $k$-nearest neighbor based Mutual Information Test (MINT, \cite{berrett2019nonparametric}) with averaging over $k$, BERET \citep{lee2023beret}, and the multiscale Fisher's independence test (MultiFIT, \cite{gorsky2018multiscale}).

\begin{figure}[h!]
	\begin{center}
		\includegraphics[width=0.975\textwidth,height=9.25cm]{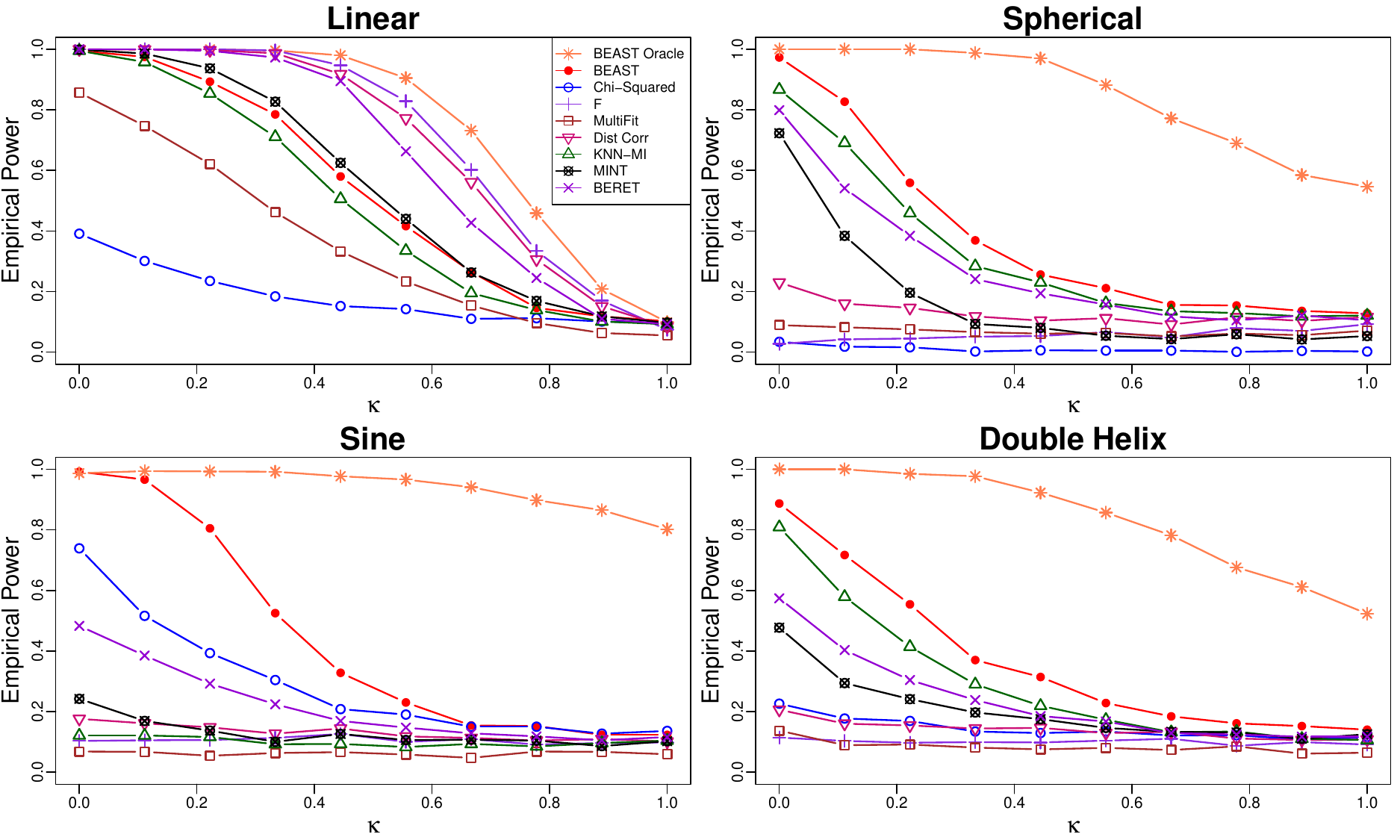}
	\end{center}
	\caption{The power curves of various methods when testing the independence between $(X_1,X_2)$ and $Y$ under four alternatives. The depth of the BEAST is $3$ and $n=128$. The level of significance is set to be $0.1$. The BEAST with oracle provides a benchmark on the feasible power for all cases. The power of the BEAST is the highest among all tests for all nonlinear forms of dependency.}\label{fig: p3d3}
\end{figure}

The data $(x_{1,i}, x_{2,i},y_i), i=1,2,\cdots, n=128$ are generated according to the settings in Table~\ref{table: p3power} below. The values of $\kappa$ are evenly spaced over $[0,1]$ to represent the strength of noise. The parameters in the scenarios are chosen such that the power curves in Figure~\ref{fig: p3d3} show a thorough comparison over different magnitude of signals.

\begin{table}
	%\resizebox{\textwidth}{!}{
	\caption{\label{table: p3power} Simulation scenarios for $p=3$: The following variables are all independent. $\epsilon_j \sim \cN(0,1)$ for $j=1,\ldots,6$; $G_j \sim \cN(0,1)$ for $j=1,2,3$; $U_j \sim \text{Unif}[0,1]$ for $j=1,2$; $\vartheta \sim \text{Unif}[-\pi,\pi]$; and $R \sim \text{Bern}(\{-1,1\},(1/2,1/2))$. $\kappa$ is evenly spaced between 0 and 1. $h(\kappa)=\sqrt{0.68+0.64\kappa-0.32\kappa^2}.$ In the sphere setting, $||\bG||=(G_1^2+G_2^2+G_3^2)^{1/2}$. In the doule helix setting, $c_0=0.4\kappa+0.5$.}
	%}
	\def\arraystretch{1}%
	\centering
	{\footnotesize	\begin{tabular}{l|l|l} \toprule
			Scenario & Generation of $(X_1,X_2)$ & Generation of $Y$ \\ \midrule
			Linear & $(X_1,X_2)\sim \cN_2(\zero,\Ib_2)$ & $Y=0.4(1-\kappa)(X_1+X_2)+h(\kappa)\epsilon_1$\\[0.5ex]
			Sphere & $ (X_1,X_2) = (G_1/||\bG||,G_2/||\bG||)$ & $\displaystyle  Y = G_3/||\bG|| + (0.7\kappa+0.3)\epsilon_2$  \\ [0.5ex]
			Sine & $(X_1,X_2)=(U_1,U_2)$	& $\displaystyle  Y =\sin\left( 4\pi(X_1+X_2)\right) + (2\kappa+0.2)\epsilon_3 $   \\ [0.5ex]
			Double Helix & $(X_1,X_2)=(R\cos\vartheta + c_0 \epsilon_4, R\sin\vartheta +c_0\epsilon_5)$ & $Y = \vartheta + c_0\epsilon_6$  \\ \bottomrule
		\end{tabular}
	}
\end{table}

The messages from Figure~\ref{fig: p3d3} are similar to those when $p=2.$ The BEAST with oracle leads the power under all scenarios to provide a benchmark for feasible power. In particular, under the ``Linear'' scenario, the gain of the power curve of $B_{\text{oracle}}$ from those of the $F$-test and the distance correlation demonstrates the ability of $B_{\text{oracle}}$ to approximate the optimal power. Similar to what we observed in the bivariate cases, the huge margin between the power curve of $B_{\text{oracle}}$ and other tests indicates the potential substantial gain in power with a proper choice of weights. By approximating the BEAST with oracle, $B_{\lambda}$ achieves robust power against any form of alternative. The BEAST is particularly powerful against complex nonlinear forms of dependency, and its power curve leads others with a huge margin under all three nonlinear scenarios.

In summary, our simulations in this section show that $B_{\lambda}$ can approximate the optimal power benchmarked by $B_{\text{oracle}}$. The BEAST demonstrates a robust power against many common alternatives in both dimensions $p=2$ or $3$. The BEAST is particularly powerful against a large class of complex nonlinear forms of dependency.

%=================================================================================================%
\section{Empirical Data Analysis}\label{sec: data}
In this section, we apply the BEAST method to the $n=256$ visually brightest stars from the Hipparcos catalog \citep{hoffleit1987bright,perryman1997hipparcos}. For each star, a number of features about its location and brightness are recorded. Here, we are interested in detecting if there exists any dependence between the joint galactic coordinates ($X_1,X_2$) and the brightness of stars. We consider the absolute magnitude in this section, while study the visual magnitude in the Supplementary Materials. We consider the BEAST, $\chi^2$-test, $F$-test, distant correlation (Dist Corr), KNN-MI, MINT, and MultiFIT to this problem. The $p$-values of all the approaches are summarized in Table \ref{tab:data:star}.
The BEAST is constructed with $m=128$, $r=48$, $\lambda=\sqrt{(pD\log 2)(8n)^{-1}}=0.055$, and $\cL=\cL_{3,D, \text{joint cross}}$ defined in (\ref{eqn:l3:cross}) where $D=3$.

\begin{table}
	\caption{\label{tab:data:star}  The $p$-values of various methods in testing the independence between the location and brightness of stars.}
	\centering
	{\footnotesize	
		\begin{tabular}{c|c|c|c|c|c|c|c}
			\toprule
			& BEAST & Dist Corr & $\chi^2$-test & $F$-test & MultiFIT & KNN-MI & MINT \\	\midrule
			$p$-value & 0 & 0 & 0.091 & 6e-5 & 0.011 & 0.34 & 0.01 \\	\bottomrule
		\end{tabular}
	}
\end{table}

When testing the independence between the absolute magnitude and the galactic coordinates, this hypothesis is significant based on all the methods except KNN-MI. In addition to producing $p$-values, the BEAST is capable to provide interpretation of the dependence while most competing methods cannot. Hence, we investigated the most important binary interaction among all possible combinations when analyzing the absolute magnitude. 
From each subsample, we record the most significant binary interaction. The most frequently occurred such interaction is 
$
\Lambda = \left( \begin{array}{ccc} 0 & 0 & 0 \\ 1 & 1 & 0 \\ 1 & 0 & 0
\end{array}\right).
$
\begin{figure}[h!]
	\centering
	\includegraphics[width=45mm, height=45mm]{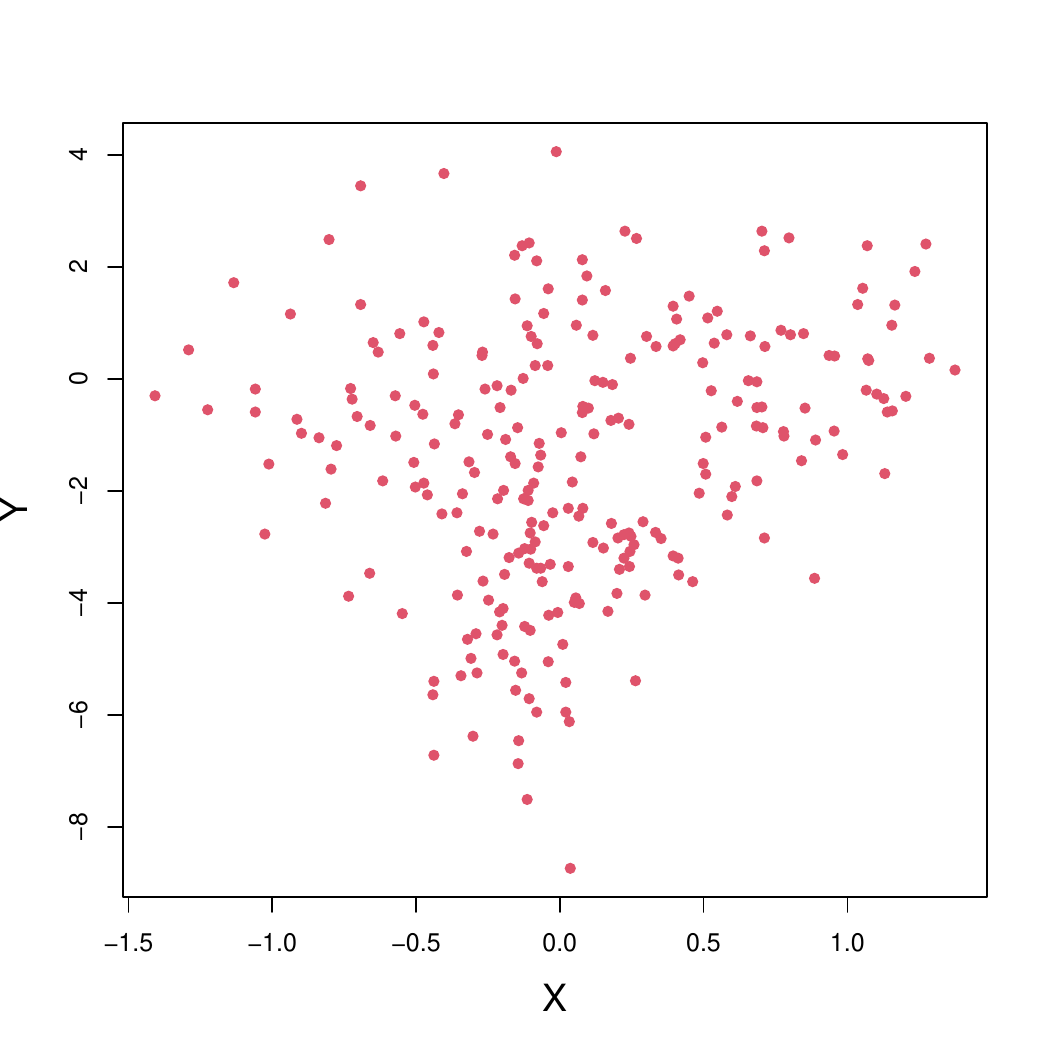}	
	\includegraphics[width=45mm, height=45mm]{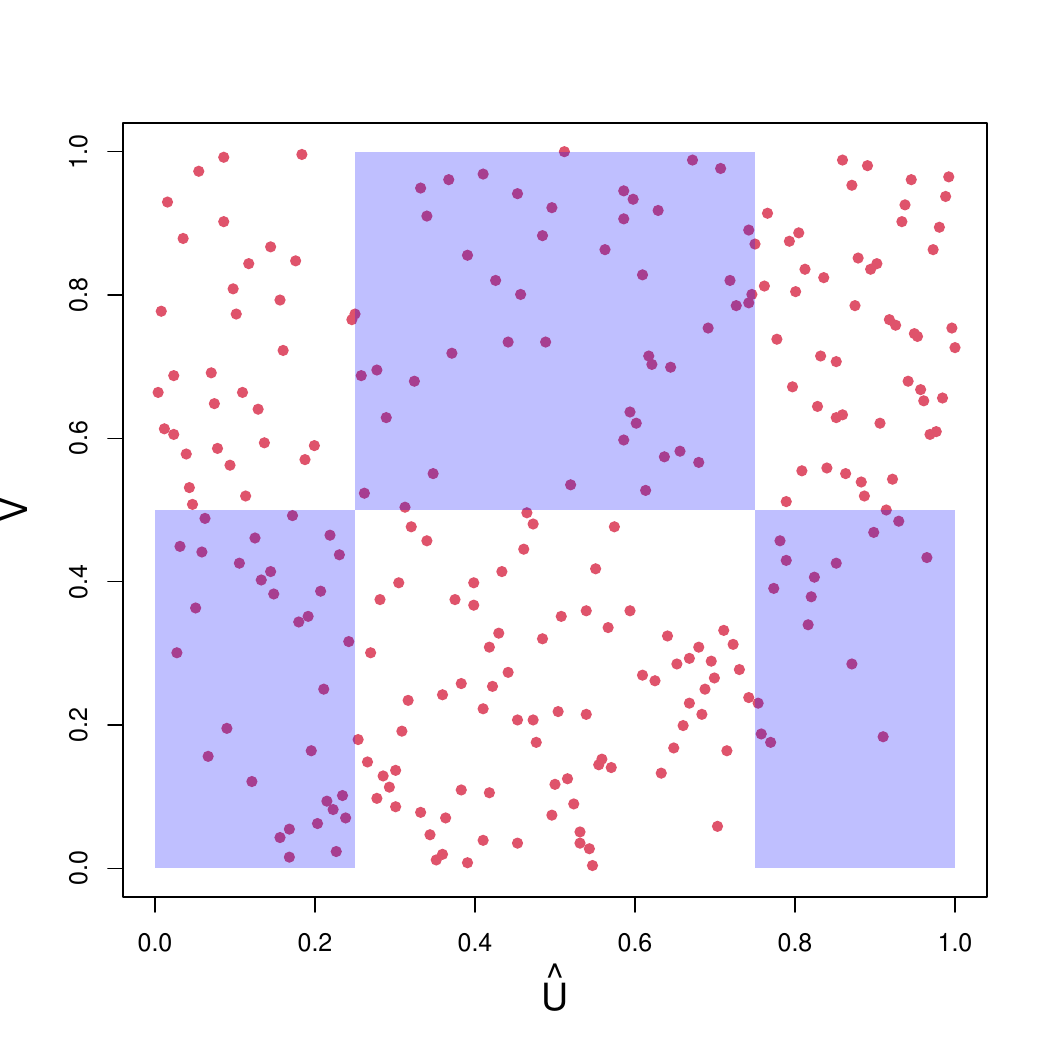}	
	\includegraphics[width=45mm, height=45mm]{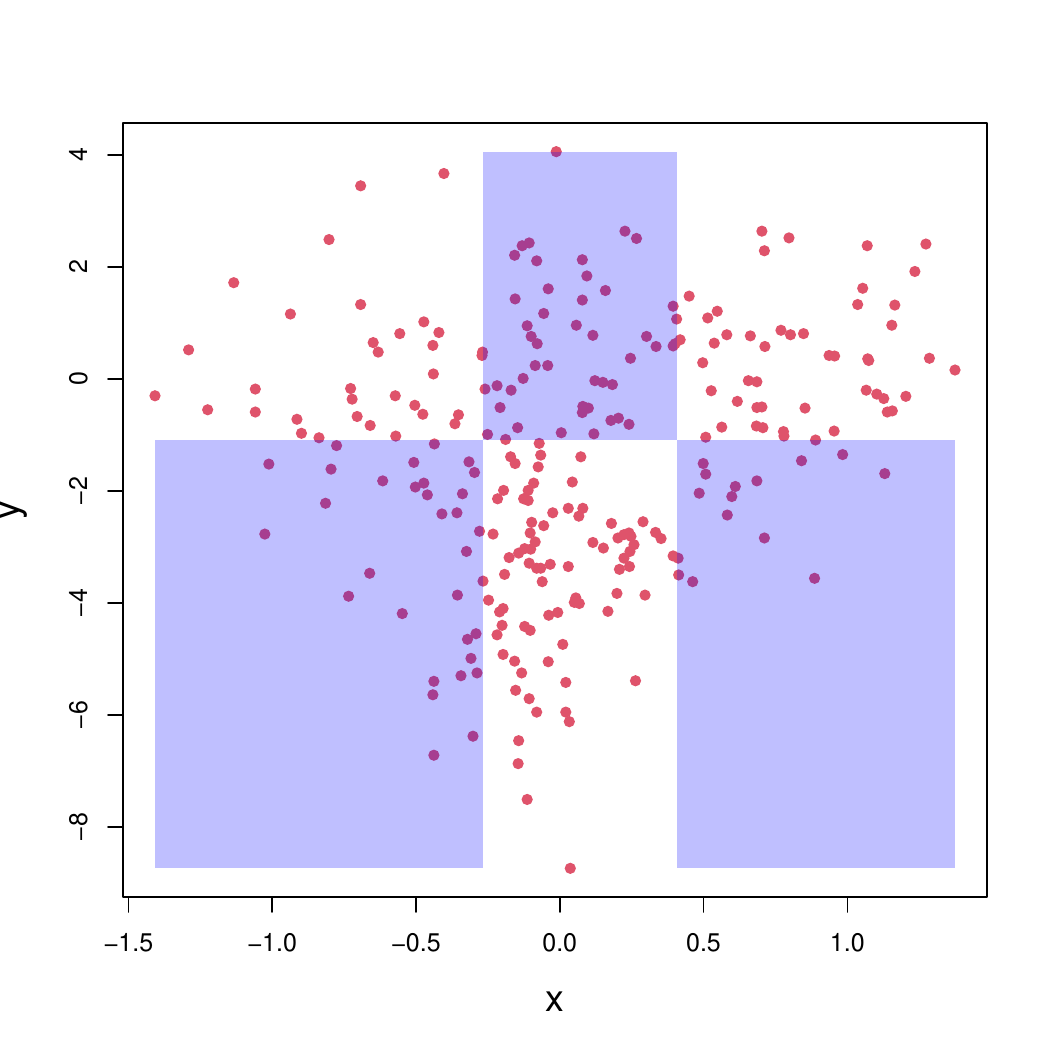}	
	\caption{\scriptsize Display of the binary interaction explaining the relationship between the location and brightness of stars. The left panel shows the scatter plot of galactic latitude ($X$) and absolute magnitude ($Y$) on the original scale. The middle panel shows the empirical copula of this distribution, equipped with the most frequent binary interaction in subsamples. {There are 162 points in white regions in contrast to 94 points in blue regions, resulting in a symmetry statistic is $68$ and a $Z$-statistic of $4.25$} for testing the balance of points in white regions and blue regions. The right panel shows the scatter plot on the original scale equipped with the same binary interaction. We notice that brighter stars (lower $Y$) tend to fall between $-16.1^\circ$ and $23.4^\circ$ in latitude, while darker stars (higher $Y$) tend to be outside this interval of $X$. This pattern provides a scientifically meaningful explanation of the statistical significance.
	}\label{fig:bid:abs}
\end{figure}
Note that for this $\Lambda$ with a first row of $0$'s, the first dimension (the galactic longitude) is not involved. In Figure \ref{fig:bid:abs}, we plot the absolute magnitude against the galactic latitude. The left panel is the scatter plot of these two variables; the middle panel is the scatter plot after the copula transformation, grouped according to the aforementioned $\Lambda$, with the white regions indicating positive interaction and blue regions indicating negative interaction; the right panel is the scatter plot on the original scale when grouped according to the same $\Lambda$. The symmetry statistic for $\Lambda$ is $68$, resulting in a $Z$-statistic of $4.25$ for testing the balance of points in white regions and blue regions. Among these stars with the most absolute magnitude, the majority of them are placed between $-16.1^\circ$ and $23.4^\circ$ in latitude. Note that in the galactic coordinate system, the fundamental plane is approximately the galactic plane of the Milky Way galaxy. Therefore, the most frequent binary interaction $\Lambda$ makes scientific sense for the statistical significance: the bright stars in the data are around the fundamental plane of the Milky Way galaxy. This clear scientific interpretation of the statistical significance is an advantage of the BEAST and the general BET framework.

%=================================================================================================%
\section{Summary and Discussions}\label{sec: disc}
We study the classical problem of nonparametric dependence detection through a novel perspective of binary expansion. The novel insights from the extension of the Euler formula and the binary expansion approximation of uniformity (BEAUTY) shed lights on the unification of important tests into the novel framework of the binary expansion adaptive symmetry test (BEAST), which considers a data-adaptively weighted sum of symmetry statistics from the binary expansion. The one-dimensional oracle on the weights leads to a benchmark of optimal power for nonparametric tests while being agnostic of the alternative. By approximating the oracle weights with resampling and regularization, the proposed BEAST demonstrates consistent performance and is effectively powerful against a variety of complex dependency forms, showcasing its potential across diverse scenarios.

%provides robust power, and is particularly powerful against a large class of complex forms of dependency.

Our study on powerful nonparametric tests of uniformity can be further extended and generalized to many directions. For example, extensions to general goodness-of-fit tests and two-sample tests can be investigated through the BEAST approach. Recent papers in these directions include \cite{brown2021august,zhao2023}. Tests of other distributional properties related to uniformity, such as tests of Gaussianity and tests of multivariate symmetry can also be studied through the BEAST approach. Insights from these tests can be used in constructing distribution-free models of dependence \cite{brown2022belief}.

With Theorem~\ref{thm: beauty}, we further observe that for two arbitrary random vectors $\bU_1$ and $\bU_2$ distributed within $[-1,1]^{p_1}$ and $[-1,1]^{p_2}$ respectively, by comparing the characteristic function of the joint distribution and the product of marginal characteristic distributions, the null hypothesis that $\bU_{1,D}$ and $\bU_{2,D}$ are independent is equivalent to the following hypothesis over the expectations of $A_\Lambda$'s:
$$
H_{0,D}: Cov(A_{\Lambda_1},A_{\Lambda_2})=0  \text{ for } \Lambda_1 \in \cL_{p_1,D,~\text{unif}} \text{ and } \Lambda_2 \in \cL_{p_2,D,~\text{unif}}.
$$ 
With this insight, we propose to study this general test of independence between random vectors in future work. In particular, we plan to study the connection between smooth forms of dependency and the order of binary interactions. A good understanding of this connection can help the development of distribution-free tests targeting at smooth forms of dependence.

Our simulation studies show a gap in empirical power between the BEAST and the BEAST with oracle. Thus the optimal trade-off between sample size, dimension, the depth of binary expansion, and the strength of the non-uniformity would be another interesting problem for investigation. The optimal subsampling and thresholding procedures are critical as well. Results on these problems would lead to a BEAST that is adaptively optimal for a  wide class of distributions in power.

\section*{Software}
The \texttt{R} function \texttt{BEAST} is freely available in the \texttt{R} package of \texttt{BET}.

\section*{Acknowledgement}
 
Zhang's research was partially supported by NSF DMS-1613112, NSF IIS-1633212, NSF DMS-1916237 and DMS-2152289. Zhao's research was partially supported by NSF IIS-1633283 and DMS-2311216.  Zhou's research was partially supported by NSF-IIS 1545994, NSF IOS-1922701, DOE DE-SC0018344. The authors thank Dr. Xiao-Li Meng and Dr. David Donoho for their generous and insightful comments that substantially improve the paper. 
% The authors also thank the editors and reviewers for very helpful comments and suggestions that substantially improved the paper. 

\bibliographystyle{abbrvnat}

\bibliography{KZ_BIB_093023}

\begin{thebibliography}{83}
\providecommand{\natexlab}[1]{#1}
\providecommand{\url}[1]{\texttt{#1}}
\expandafter\ifx\csname urlstyle\endcsname\relax
  \providecommand{\doi}[1]{doi: #1}\else
  \providecommand{\doi}{doi: \begingroup \urlstyle{rm}\Url}\fi

\bibitem[Anderson and Darling(1954)]{anderson1954test}
T.~W. Anderson and D.~A. Darling.
\newblock A test of goodness of fit.
\newblock \emph{Journal of the American statistical association}, 49\penalty0
  (268):\penalty0 765--769, 1954.

\bibitem[Balakrishnan and
  Wasserman(2019{\natexlab{a}})]{Balakrishnan2019hypothesis}
S.~Balakrishnan and L.~Wasserman.
\newblock {Hypothesis testing for densities and high-dimensional multinomials:
  Sharp local minimax rates}.
\newblock \emph{The Annals of Statistics}, 47\penalty0 (4):\penalty0 1893 --
  1927, 2019{\natexlab{a}}.
\newblock \doi{10.1214/18-AOS1729}.
\newblock URL \url{https://doi.org/10.1214/18-AOS1729}.

\bibitem[Balakrishnan and Wasserman(2019{\natexlab{b}})]{wasserman:2019}
S.~Balakrishnan and L.~Wasserman.
\newblock {Hypothesis testing for densities and high-dimensional multinomials:
  Sharp local minimax rates}.
\newblock \emph{The Annals of Statistics}, 47\penalty0 (4):\penalty0 1893 --
  1927, 2019{\natexlab{b}}.
\newblock \doi{10.1214/18-AOS1729}.
\newblock URL \url{https://doi.org/10.1214/18-AOS1729}.

\bibitem[Barber and Cand\`{e}s(2019)]{BarberandCandes2019}
R.~F. Barber and E.~J. Cand\`{e}s.
\newblock A knockoff filter for high-dimensional selective inference.
\newblock \emph{The Annals of Statistics}, 47\penalty0 (5):\penalty0
  2504--2537, 2019.

\bibitem[Barber and Janson(2022)]{barber2022testing}
R.~F. Barber and L.~Janson.
\newblock Testing goodness-of-fit and conditional independence with approximate
  co-sufficient sampling.
\newblock \emph{The Annals of Statistics}, 50\penalty0 (5):\penalty0
  2514--2544, 2022.

\bibitem[Barron(1989)]{barron1989uniformly}
A.~R. Barron.
\newblock Uniformly powerful goodness of fit tests.
\newblock \emph{The Annals of Statistics}, pages 107--124, 1989.

\bibitem[Berrett and Samworth(2019)]{berrett2019nonparametric}
T.~B. Berrett and R.~J. Samworth.
\newblock Nonparametric independence testing via mutual information.
\newblock \emph{Biometrika}, 106\penalty0 (3):\penalty0 547--566, 2019.

\bibitem[Berrett and Samworth(2021)]{berrett2021usp}
T.~B. Berrett and R.~J. Samworth.
\newblock Usp: an independence test that improves on pearson’s chi-squared
  and the g-test.
\newblock \emph{Proceedings of the Royal Society A}, 477\penalty0
  (2256):\penalty0 20210549, 2021.

\bibitem[Berrett et~al.(2021)Berrett, Kontoyiannis, and
  Samworth]{berrett2021optimal}
T.~B. Berrett, I.~Kontoyiannis, and R.~J. Samworth.
\newblock Optimal rates for independence testing via u-statistic permutation
  tests.
\newblock \emph{The Annals of Statistics}, 49\penalty0 (5):\penalty0
  2457--2490, 2021.

\bibitem[Blum et~al.(1961)Blum, Kiefer, and Rosenblatt]{blum1961}
J.~R. Blum, J.~Kiefer, and M.~Rosenblatt.
\newblock {Distribution Free Tests of Independence Based on the Sample
  Distribution Function}.
\newblock \emph{The Annals of Mathematical Statistics}, 32\penalty0
  (2):\penalty0 485 -- 498, 1961.
\newblock \doi{10.1214/aoms/1177705055}.
\newblock URL \url{https://doi.org/10.1214/aoms/1177705055}.

\bibitem[Box et~al.(2005)Box, Hunter, and Hunter]{box2005statistics}
G.~E. Box, J.~S. Hunter, and W.~G. Hunter.
\newblock \emph{Statistics for experimenters: design, innovation, and
  discovery}, volume~2.
\newblock Wiley-Interscience New York, 2005.

\bibitem[Brown and Zhang(2023)]{brown2021august}
B.~Brown and K.~Zhang.
\newblock The {AUGUST} two-sample test: Powerful, interpretable, and fast.
\newblock \emph{The New England Journal of Statistics in Data Science},
  \penalty0 (accepted), 2023.

\bibitem[Brown et~al.(2022)Brown, Zhang, and Meng]{brown2022belief}
B.~Brown, K.~Zhang, and X.-L. Meng.
\newblock Belief in dependence: Leveraging atomic linearity in data bits for
  rethinking generalized linear models.
\newblock \emph{arXiv preprint arXiv:2210.10852}, 2022.

\bibitem[Cao and Bickel(2020)]{cao2020correlations}
S.~Cao and P.~J. Bickel.
\newblock Correlations with tailored extremal properties.
\newblock \emph{arXiv preprint arXiv:2008.10177}, 2020.

\bibitem[Chatterjee(2020)]{chatterjee2019new}
S.~Chatterjee.
\newblock A new coefficient of correlation.
\newblock \emph{Journal of the American Statistical Association}, page in
  press, 2020.

\bibitem[Cox(1975)]{Cox1975}
D.~Cox.
\newblock A note on data-splitting for the evaluation of significance levels.
\newblock \emph{Biometrika}, 62\penalty0 (2):\penalty0 441--444, 1975.

\bibitem[Cox and Reid(2000)]{cox2000theory}
D.~R. Cox and N.~Reid.
\newblock \emph{The theory of the design of experiments}.
\newblock CRC Press, 2000.

\bibitem[Cressie and Read(1984)]{cressie1984multinomial}
N.~Cressie and T.~R. Read.
\newblock Multinomial goodness-of-fit tests.
\newblock \emph{Journal of the Royal Statistical Society: Series B
  (Methodological)}, 46\penalty0 (3):\penalty0 440--464, 1984.

\bibitem[Dai et~al.(2020)Dai, Lin, Xing, and Liu]{DaiLinXingLiu2020}
C.~Dai, B.~Lin, X.~Xing, and J.~Liu.
\newblock False discovery rate control via data splitting.
\newblock \emph{arXiv preprint arXiv:2002.08542v1}, 2020.

\bibitem[Deb et~al.(2020)Deb, Ghosal, and Sen]{deb2020measuring}
N.~Deb, P.~Ghosal, and B.~Sen.
\newblock Measuring association on topological spaces using kernels and
  geometric graphs.
\newblock \emph{arXiv preprint arXiv:2010.01768}, 2020.

\bibitem[Diaz~Rivero and Dvorkin(2020)]{rivero2020flow}
A.~Diaz~Rivero and C.~Dvorkin.
\newblock Flow-based likelihoods for non-gaussian inference.
\newblock \emph{Phys. Rev. D}, 102:\penalty0 103507, Nov 2020.
\newblock \doi{10.1103/PhysRevD.102.103507}.
\newblock URL \url{https://link.aps.org/doi/10.1103/PhysRevD.102.103507}.

\bibitem[Donoho and Johnstone(1994)]{donoho1994ideal}
D.~L. Donoho and J.~M. Johnstone.
\newblock Ideal spatial adaptation by wavelet shrinkage.
\newblock \emph{Biometrika}, 81\penalty0 (3):\penalty0 425--455, 1994.

\bibitem[Efron and Tibshirani(1994)]{efron1994introduction}
B.~Efron and R.~J. Tibshirani.
\newblock \emph{An introduction to the bootstrap}.
\newblock CRC press, 1994.

\bibitem[Escanciano(2006)]{escanciano2006goodness}
J.~C. Escanciano.
\newblock Goodness-of-fit tests for linear and nonlinear time series models.
\newblock \emph{Journal of the American Statistical Association}, 101\penalty0
  (474):\penalty0 531--541, 2006.

\bibitem[Fan and Huang(2001)]{fan2001goodness}
J.~Fan and L.-S. Huang.
\newblock Goodness-of-fit tests for parametric regression models.
\newblock \emph{Journal of the American Statistical Association}, 96\penalty0
  (454):\penalty0 640--652, 2001.

\bibitem[Fan et~al.(2019)Fan, Ke, Sun, and Zhou]{FARM2019}
J.~Fan, Y.~Ke, Q.~Sun, and W.~Zhou.
\newblock Farmtest: Factor-adjusted robust multiple testing with approximate
  false discovery control.
\newblock \emph{Journal of the American Statistical Association}, 114\penalty0
  (528):\penalty0 1880--1893, 2019.

\bibitem[Geenens and de~Micheaux(2020)]{Geenens2021new}
G.~Geenens and P.~de~Micheaux.
\newblock The hellinger correlation.
\newblock \emph{Journal of the American Statistical Association}, page in
  press, 2020.

\bibitem[Genest and Verret(2005)]{genest2005locally}
C.~Genest and F.~Verret.
\newblock Locally most powerful rank tests of independence for copula models.
\newblock \emph{Nonparametric Statistics}, 17\penalty0 (5):\penalty0 521--539,
  2005.

\bibitem[Genest et~al.(2006)Genest, Quessy, and
  R{\'e}millard]{genest2006goodness}
C.~Genest, J.-F. Quessy, and B.~R{\'e}millard.
\newblock Goodness-of-fit procedures for copula models based on the probability
  integral transformation.
\newblock \emph{Scandinavian Journal of Statistics}, 33\penalty0 (2):\penalty0
  337--366, 2006.

\bibitem[Genest et~al.(2009)Genest, R{\'e}millard, and
  Beaudoin]{genest2009goodness}
C.~Genest, B.~R{\'e}millard, and D.~Beaudoin.
\newblock Goodness-of-fit tests for copulas: A review and a power study.
\newblock \emph{Insurance: Mathematics and economics}, 44\penalty0
  (2):\penalty0 199--213, 2009.

\bibitem[Genest et~al.(2019)Genest, Ne{\v{s}}lehov{\'a}, R{\'e}millard, and
  Murphy]{genest2019testing}
C.~Genest, J.~Ne{\v{s}}lehov{\'a}, B.~R{\'e}millard, and O.~Murphy.
\newblock Testing for independence in arbitrary distributions.
\newblock \emph{Biometrika}, 106\penalty0 (1):\penalty0 47--68, 2019.

\bibitem[Golubov et~al.(2012)Golubov, Efimov, and Skvortsov]{golubov2012walsh}
B.~Golubov, A.~Efimov, and V.~Skvortsov.
\newblock \emph{Walsh series and transforms: theory and applications},
  volume~64.
\newblock Springer Science \& Business Media, 2012.

\bibitem[Gonz{\'a}lez-Manteiga and Crujeiras(2013)]{gonzalez2013updated}
W.~Gonz{\'a}lez-Manteiga and R.~M. Crujeiras.
\newblock An updated review of goodness-of-fit tests for regression models.
\newblock \emph{Test}, 22:\penalty0 361--411, 2013.

\bibitem[Gorsky and Ma(2018)]{gorsky2018multiscale}
S.~Gorsky and L.~Ma.
\newblock Multiscale $\text{Fisher's}$ independence test for multivariate
  dependence.
\newblock \emph{arXiv preprint arXiv:1806.06777}, 2018.

\bibitem[Gretton et~al.(2007)Gretton, Fukumizu, Teo, Song, Sch{\"o}lkopf, and
  Smola]{gretton2007kernel}
A.~Gretton, K.~Fukumizu, C.~H. Teo, L.~Song, B.~Sch{\"o}lkopf, and A.~J. Smola.
\newblock A kernel statistical test of independence.
\newblock In \emph{Advances in neural information processing systems}, pages
  585--592, 2007.

\bibitem[Harmuth(2013)]{harmuth2013transmission}
H.~Harmuth.
\newblock \emph{Transmission of Information by Orthogonal Functions}.
\newblock Springer Berlin Heidelberg, 2013.
\newblock ISBN 9783662132272.
\newblock URL \url{https://books.google.com/books?id=O67wCAAAQBAJ}.

\bibitem[Hartigan(1969)]{Hartigan1969}
J.~Hartigan.
\newblock Using subsample values as typical values.
\newblock \emph{Journal of the American Statistical Association}, 64\penalty0
  (328):\penalty0 1303--1317, 1969.

\bibitem[Heller and Heller(2016)]{heller2016multivariate}
R.~Heller and Y.~Heller.
\newblock Multivariate tests of association based on univariate tests.
\newblock In D.~D. Lee, M.~Sugiyama, U.~V. Luxburg, I.~Guyon, and R.~Garnett,
  editors, \emph{Advances in Neural Information Processing Systems 29}, pages
  208--216. Curran Associates, Inc., 2016.
\newblock URL
  \url{http://papers.nips.cc/paper/6220-multivariate-tests-of-association-based-on-univariate-tests.pdf}.

\bibitem[Heller et~al.(2013)Heller, Heller, and Gorfine]{heller2012consistent}
R.~Heller, Y.~Heller, and M.~Gorfine.
\newblock A consistent multivariate test of association based on ranks of
  distances.
\newblock \emph{Biometrika}, 100\penalty0 (2):\penalty0 503--510, 2013.

\bibitem[Heller et~al.(2016)Heller, Heller, Kaufman, Brill, and
  Gorfine]{heller2016consistent}
R.~Heller, Y.~Heller, S.~Kaufman, B.~Brill, and M.~Gorfine.
\newblock Consistent distribution-free $ k $-sample and independence tests for
  univariate random variables.
\newblock \emph{Journal of Machine Learning Research}, 17\penalty0
  (29):\penalty0 1--54, 2016.

\bibitem[Hoeffding(1948)]{hoeffding1948non}
W.~Hoeffding.
\newblock A non-parametric test of independence.
\newblock \emph{The Annals of Mathematical Statistics}, pages 546--557, 1948.

\bibitem[Hoffleit and Warren~Jr(1987)]{hoffleit1987bright}
D.~Hoffleit and W.~Warren~Jr.
\newblock The bright star catalogue.
\newblock \emph{ADCBu}, 1\penalty0 (4), 1987.

\bibitem[Huang(2015)]{Huang2015}
Y.~Huang.
\newblock \emph{Projection test for high-dimensional mean vectors with optimal
  direction}.
\newblock PhD thesis, Dept. of Stat., The Pennsylvania State University, 2015.

\bibitem[Ignatiadis et~al.(2016)Ignatiadis, Klaus, Zaugg, and
  Huber]{Ignatiadis2016}
N.~Ignatiadis, B.~Klaus, J.~Zaugg, and W.~Huber.
\newblock Data-driven hypothesis weighting increases detection power in
  genome-scale multiple testing.
\newblock \emph{Nature Methods}, 13\penalty0 (7):\penalty0 577--580, 2016.

\bibitem[Jager and Wellner(2007)]{jager2007goodness}
L.~Jager and J.~A. Wellner.
\newblock Goodness-of-fit tests via phi-divergences.
\newblock \emph{Annals of Statistics}, 35\penalty0 (5):\penalty0 2018--2053,
  2007.

\bibitem[Jankov{\'a} et~al.(2020)Jankov{\'a}, Shah, B{\"u}hlmann, and
  Samworth]{jankova2020goodness}
J.~Jankov{\'a}, R.~D. Shah, P.~B{\"u}hlmann, and R.~J. Samworth.
\newblock Goodness-of-fit testing in high dimensional generalized linear
  models.
\newblock \emph{Journal of the Royal Statistical Society Series B: Statistical
  Methodology}, 82\penalty0 (3):\penalty0 773--795, 2020.

\bibitem[Jin and Matteson(2018)]{jinmatteson18}
Z.~Jin and D.~Matteson.
\newblock Generalizing distance covariance to measure and test multivariate
  mutual dependence via complete and incomplete v-statistics.
\newblock \emph{Journal of Multivariate Analysis}, 168:\penalty0 304--322,
  2018.

\bibitem[Kac(1959)]{kac1959statistical}
M.~Kac.
\newblock \emph{Statistical independence in probability, analysis and number
  theory}, volume 134.
\newblock Mathematical Association of America, 1959.

\bibitem[Kaiser and Soumendra(2012)]{kaiser2012goodness}
M.~S. Kaiser and N.~Soumendra.
\newblock Goodness of fit tests for a class of markov random field models.
\newblock \emph{The Annals of Statistics}, 40\penalty0 (1):\penalty0 104--130,
  2012.

\bibitem[Kim and Ramdas(2020)]{kim2020dimension}
I.~Kim and A.~Ramdas.
\newblock Dimension-agnostic inference.
\newblock \emph{arXiv preprint arXiv:2011.05068}, 2020.

\bibitem[Kinney and Atwal(2014)]{kinney2014equitability}
J.~Kinney and G.~Atwal.
\newblock Equitability, mutual information, and the maximal information
  coefficient.
\newblock \emph{Proceedings of the National Academy of Sciences}, 111\penalty0
  (9):\penalty0 3354--3359, 2014.

\bibitem[Kojadinovic and Holmes(2009)]{kojadinovic2009tests}
I.~Kojadinovic and M.~Holmes.
\newblock Tests of independence among continuous random vectors based on
  cram{\'e}r--von mises functionals of the empirical copula process.
\newblock \emph{Journal of Multivariate Analysis}, 100\penalty0 (6):\penalty0
  1137--1154, 2009.

\bibitem[LeCam(1973)]{lecam1973convergence}
L.~LeCam.
\newblock Convergence of estimates under dimensionality restrictions.
\newblock \emph{The Annals of Statistics}, pages 38--53, 1973.

\bibitem[Lee et~al.(2023)Lee, Zhang, and Kosorok]{lee2023beret}
D.~Lee, K.~Zhang, and M.~R. Kosorok.
\newblock The binary expansion randomized ensemble test.
\newblock \emph{Statistica Sinica}, 33\penalty0 (4), 2023.

\bibitem[Lehmann and Romano(2006)]{lehmann2006testing}
E.~L. Lehmann and J.~P. Romano.
\newblock \emph{Testing statistical hypotheses}.
\newblock Springer Science \& Business Media, 2006.

\bibitem[Liang et~al.(2001)Liang, Fang, Hickernell, and Li]{liang2001testing}
J.-J. Liang, K.-T. Fang, F.~Hickernell, and R.~Li.
\newblock Testing multivariate uniformity and its applications.
\newblock \emph{Mathematics of Computation}, 70\penalty0 (233):\penalty0
  337--355, 2001.

\bibitem[Liu et~al.(2019)Liu, Yu, and Li]{LiuYuLi2020}
W.~Liu, X.~Yu, and R.~Li.
\newblock Multiple-splitting projection test for high-dimensional mean vectors.
\newblock \emph{Dept. of Stat., Stanford University}, 2019.
\newblock Unpublished.

\bibitem[Lynn(1973)]{lynn1973introduction}
P.~A. Lynn.
\newblock \emph{An introduction to the analysis and processing of signals}.
\newblock McMillan, 1973.

\bibitem[Ma and Mao(2019)]{ma2019fisher}
L.~Ma and J.~Mao.
\newblock Fisher exact scanning for dependency.
\newblock \emph{Journal of the American Statistical Association}, 114\penalty0
  (525):\penalty0 245--258, 2019.

\bibitem[Miller and Siegmund(1982)]{miller1982maximally}
R.~Miller and D.~Siegmund.
\newblock Maximally selected chi square statistics.
\newblock \emph{Biometrics}, 38\penalty0 (4):\penalty0 1011--1016, 1982.

\bibitem[Nelsen(2007)]{nelsen2007introduction}
R.~B. Nelsen.
\newblock \emph{An introduction to copulas}.
\newblock Springer Science \& Business Media, 2007.

\bibitem[Neyman and Pearson(1933)]{neyman1933ix}
J.~Neyman and E.~Pearson.
\newblock Ix. on the problem of the most efficient tests of statistical
  hypotheses.
\newblock \emph{Philosophical Transactions of the Royal Society of London.
  Series A}, 231:\penalty0 289--337, 1933.

\bibitem[Pearl(1971)]{Pearl1971}
J.~Pearl.
\newblock Application of walsh transform to statistical analysis.
\newblock \emph{IEEE Transactions on Systems, Man, and Cybernetics},
  SMC-1\penalty0 (2):\penalty0 111--119, 1971.
\newblock ISSN 0018-9472.
\newblock \doi{10.1109/TSMC.1971.4308267}.

\bibitem[Perryman et~al.(1997)Perryman, Lindegren, Kovalevsky, Hoeg, Bastian,
  Bernacca, Cr{\'e}z{\'e}, Donati, Grenon, Grewing, and
  Van~Leeuwen]{perryman1997hipparcos}
M.~Perryman, L.~Lindegren, J.~Kovalevsky, E.~Hoeg, U.~Bastian, P.~Bernacca,
  M.~Cr{\'e}z{\'e}, F.~Donati, M.~Grenon, M.~Grewing, and F.~Van~Leeuwen.
\newblock The hipparcos catalogue.
\newblock \emph{Astronomy and Astrophysics}, 323\penalty0 (1):\penalty0 49--52,
  1997.

\bibitem[Pfister et~al.(2016)Pfister, B{\"u}hlmann, Sch{\"o}lkopf, and
  Peters]{pfister2016kernel}
N.~Pfister, P.~B{\"u}hlmann, B.~Sch{\"o}lkopf, and J.~Peters.
\newblock Kernel-based tests for joint independence.
\newblock \emph{arXiv preprint arXiv:1603.00285}, 2016.

\bibitem[Politis et~al.(1999)Politis, Romano, and Wolf]{Politis.etal1999}
D.~N. Politis, J.~P. Romano, and M.~Wolf.
\newblock \emph{Subsampling}.
\newblock Springer, New York, 1999.
\newblock ISBN 0387988548 9780387988542.
\newblock URL
  \url{http://www.worldcat.org/search?qt=worldcat_org_all&q=9780387988542}.

\bibitem[Reshef et~al.(2011)Reshef, Reshef, Finucane, Grossman, McVean,
  Turnbaugh, Lander, Mitzenmacher, and Sabeti]{reshef2011detecting}
D.~N. Reshef, Y.~A. Reshef, H.~K. Finucane, S.~R. Grossman, G.~McVean, P.~J.
  Turnbaugh, E.~S. Lander, M.~Mitzenmacher, and P.~C. Sabeti.
\newblock Detecting novel associations in large data sets.
\newblock \emph{Science}, 334\penalty0 (6062):\penalty0 1518--1524, 2011.

\bibitem[Romano and DiCiccio(2019)]{RomanoDiCiccio2019}
J.~Romano and C.~DiCiccio.
\newblock Multiple data splitting for testing.
\newblock \emph{Department of Statistics, Stanford University}, 2019.
\newblock Unpublished manuscript.

\bibitem[Sejdinovic et~al.(2014)Sejdinovic, Sriperumbudur, Gretton, and
  Fukumizu]{sejdinovic2013}
D.~Sejdinovic, B.~Sriperumbudur, A.~Gretton, and K.~Fukumizu.
\newblock Equivalence of distance-based and $\text{RKHS}$-based statistics in
  hypothesis testing.
\newblock \emph{The Annals of Statistics}, 10\penalty0 (5):\penalty0
  2263--2291, 2014.

\bibitem[Sen and Sen(2014)]{sen2014testing}
A.~Sen and B.~Sen.
\newblock Testing independence and goodness-of-fit in linear models.
\newblock \emph{Biometrika}, 101\penalty0 (4):\penalty0 927--942, 2014.

\bibitem[Shi et~al.(2020)Shi, Hallin, Drton, and Han]{shi2020rate}
H.~Shi, M.~Hallin, M.~Drton, and F.~Han.
\newblock Rate-optimality of consistent distribution-free tests of independence
  based on center-outward ranks and signs.
\newblock \emph{arXiv preprint arXiv:2007.02186}, 2020.

\bibitem[Spearman(1904)]{spearman1904proof}
C.~Spearman.
\newblock The proof and measurement of association between two things.
\newblock \emph{The American Journal of Psychology}, 15\penalty0 (1):\penalty0
  72--101, 1904.

\bibitem[Stephens(1976)]{stephens1976asymptotic}
M.~A. Stephens.
\newblock Asymptotic results for goodness-of-fit statistics with unknown
  parameters.
\newblock \emph{The Annals of Statistics}, 4\penalty0 (2):\penalty0 357--369,
  1976.

\bibitem[Sz\'{e}kely et~al.(2007)Sz\'{e}kely, Rizzo, and Bakirov]{szekely2007}
G.~J. Sz\'{e}kely, M.~L. Rizzo, and N.~K. Bakirov.
\newblock Measuring and testing dependence by correlation of distances.
\newblock \emph{The Annals of Statistics}, 35\penalty0 (6):\penalty0
  2769--2794, 12 2007.
\newblock \doi{10.1214/009053607000000505}.
\newblock URL \url{http://dx.doi.org/10.1214/009053607000000505}.

\bibitem[Wasserman(2006)]{wasserman2006all}
L.~Wasserman.
\newblock \emph{All of Nonparametric Statistics}.
\newblock Springer Texts in Statistics. Springer New York, 2006.
\newblock ISBN 9780387306230.
\newblock URL \url{https://books.google.com/books?id=MRFlzQfRg7UC}.

\bibitem[Wasserman and Roeder(2009)]{WassermanandRoeder2009}
L.~Wasserman and K.~Roeder.
\newblock High dimensional variable selection.
\newblock \emph{The Annals of Statistics}, 37\penalty0 (5A):\penalty0
  2178--2201, 2009.

\bibitem[Xiang et~al.(2022)Xiang, Zhang, Liu, Perou, Zhang, and
  Marron]{xiang2022pairwise}
S.~Xiang, W.~Zhang, S.~Liu, C.~M. Perou, K.~Zhang, and J.~Marron.
\newblock Pairwise nonlinear dependence analysis of genomic data.
\newblock \emph{Annals of Applied Statistics}, 2022.

\bibitem[Zhang(2002)]{zhang2002powerful}
J.~Zhang.
\newblock Powerful goodness-of-fit tests based on the likelihood ratio.
\newblock \emph{Journal of the Royal Statistical Society: Series B (Statistical
  Methodology)}, 64\penalty0 (2):\penalty0 281--294, 2002.

\bibitem[Zhang(2019)]{zhang2019bet}
K.~Zhang.
\newblock {BET} on independence.
\newblock \emph{Journal of the American Statistical Association}, 114\penalty0
  (528):\penalty0 1620--1637, 2019.
\newblock \doi{10.1080/01621459.2018.1537921}.
\newblock URL \url{https://doi.org/10.1080/01621459.2018.1537921}.

\bibitem[Zhao et~al.(2023{\natexlab{a}})Zhao, Zhang, and Zhou]{zhao2023}
Z.~Zhao, K.~Zhang, and W.~Zhou.
\newblock Bit by bit: The universal binary coding of random variables in
  statistics.
\newblock 2023{\natexlab{a}}.
\newblock Technical report.

\bibitem[Zhao et~al.(2023{\natexlab{b}})Zhao, Zhang, Baiocchi, Li, and
  Zhang]{zhao2023fast}
Z.~Zhao, W.~Zhang, M.~Baiocchi, Y.~Li, and K.~Zhang.
\newblock Fast, flexible, and powerful: Introducing a scalable, bitwise
  framework for non-parametric testing for dependence structure.
\newblock \emph{Techinical report}, 2023{\natexlab{b}}.

\bibitem[Zheng et~al.(2012)Zheng, Shi, and Zhang]{zheng2012generalized}
S.~Zheng, N.-Z. Shi, and Z.~Zhang.
\newblock Generalized measures of correlation for asymmetry, nonlinearity, and
  beyond.
\newblock \emph{Journal of the American Statistical Association}, 107\penalty0
  (499):\penalty0 1239--1252, 2012.

\bibitem[Zhu et~al.(2017)Zhu, Xu, Li, and Zhong]{Zhu2017}
L.~Zhu, K.~Xu, R.~Li, and W.~Zhong.
\newblock Projection correlation between two random vectors.
\newblock \emph{Biometrika}, 104\penalty0 (4):\penalty0 829--843, 2017.

\end{thebibliography}

\end{document}